\documentclass[structabstract]{aa}  
\usepackage{natbib}
\usepackage{graphicx,breqn,color}
\usepackage{txfonts}
\usepackage{ulem}
{}

\newcommand{\myr}{M$_{\odot}$\,yr$^{-1}$}

\begin{document}
   \title{Binary evolution using the theory of osculating orbits}

   \subtitle{I. Conservative Algol evolution}

  \author{P.~J. Davis
          \inst{1}
          \and
          L. Siess\inst{1}
          \and
          R. Deschamps\inst{1,2}
          }

   \institute{Institut d'Astronomie et d'Astrophysique (IAA),
     Universit\'{e} Libre de Bruxelles (ULB), CP226, Boulevard du Triomphe,
     B-1050 Brussels, Belgium \\
              \email{pdavis@ulb.ac.be}
   \and
             ESO Vitacura, Avenue Alonso de C\'{o}rdova 3107, Vitacura,
             Casilla 19001, Santiago de
             Chile, Chile\\
   }

   \date{}

 
  \abstract
  {Studies of conservative mass transfer in interacting binary systems
    widely assume that orbital angular momentum is conserved. However, this
    only holds under physically unrealistic assumptions.}
  {Our aim is to calculate the evolution of Algol binaries within the
    framework of the osculating orbital theory, which considers the perturbing
    forces acting on the orbit of each star arising from mass exchange via
    Roche lobe overflow (RLOF). The scheme is compared to results
    calculated from a `classical' prescription.}
  {Using our stellar binary evolution code BINSTAR, we calculate the
    orbital evolution of Algol binaries undergoing case A and case B mass
    transfer, by applying the osculating scheme. The velocities of the
    ejected and accreted material are evaluated by solving the restricted
    three-body equations of motion, within the ballistic
    approximation. This allows us to determine the change of linear momentum
    of each star, and the gravitational
    force applied by the mass transfer stream. Torques applied onto the
    stellar spins by tides and mass transfer are also considered.}
  {Using the osculating formalism gives
  shorter post-mass transfer orbital periods typically by a factor of 4
  compared to the classical scheme, owing to the gravitational force
  applied onto the stars by the mass transfer stream. Additionally, during
  the rapid phase of mass exchange, the donor star is spun down on a
  timescale shorter than the tidal synchronization timescale, leading to
  sub-synchronous rotation. Consequently, between 15 and 20 per cent
  of the material leaving the inner-Lagrangian point is accreted back onto
  the donor (so-called `self-accretion'), further enhancing orbital
  shrinkage. Self-accretion, and the sink of orbital angular momentum which
  mass transfer provides, may potentially lead to more contact
  binaries. Even though Algols are mainly considered, the osculating
  prescription is applicable to all types of interacting binaries,
  including those with eccentric orbits.}
   {}


   \maketitle
%

\section{Introduction} \label{sec:intro}

Roche lobe overflow (RLOF) gives rise to a wide variety of phenomena, with
implications for many areas of astrophysics. For instance, accretion onto a
massive white dwarf may lead to Type Ia supernovae \citep[][for a
review]{1960ApJ...132..565H,2012NewAR..56..122W}, or X-ray emission from
accreting neutron stars and black holes
\citep[e.g.][]{1966ApJ...144..840Z}. Additionally, RLOF is a viable
formation channel for blue stragglers
\citep{1964MNRAS.128..147M,2011Natur.478..356G,2013MNRAS.428..897L},
sub-dwarf B stars
\citep{1976ApJ...204..488M,2002MNRAS.336..449H,2013MNRAS.tmp.1615C}, and
for interacting compact binaries via the common envelope phase, which is
triggered by dynamically unstable RLOF from a giant or asymptotic giant
branch star \citep{1976IAUS...73...75P,2008ASSL..352..233W}.



The exchange of mass and angular momentum during RLOF causes the orbital
separation to change, dictating the subsequent fate of the binary
system. For a primary (i.e. the initially more massive) star of mass
$M_{1}$ and a secondary star of mass $M_{2}$, the rate of change of the
semi-major axis, $\dot{a}$, is determined from the rates of change of the
orbital angular momentum, $\dot{J}_{\mathrm{orb}}$, the primary mass,
$\dot{M}_{1}$, the secondary mass, $\dot{M}_{2}$, and of the
eccentricity, $\dot{e}$, according to

\begin{equation}
  \frac{\dot{a}}{a}=2\frac{\dot{J}_{\mathrm{orb}}}{J_{\mathrm{orb}}}-2\frac{\dot{M}_{1}}{M_{1}}-2\frac{\dot{M}_{2}}{M_{2}}+\frac{\dot{M}_{1}+\dot{M}_{2}}{M_{1}+M_{2}}+\frac{2e\dot{e}}{1-e^{2}}.
\label{adot_a_canonical} 
\end{equation} 
For conservative mass transfer within circular orbits ($e=0$), it is
assumed that mass and orbital angular momentum are conserved, i.e.
$\dot{M}_{2}=-\dot{M}_{1}$, and $\dot{J}_{\mathrm{orb}}=0$
\citep[e.g.][]{1971ARA&A...9..183P}. Here, mass transfer from the more
massive primary to the less massive secondary ($q=M_{1}/M_{2}>1$) leads to
orbital shrinkage ($\dot{a}<0$), while the reverse occurs for $q<1$
\citep[see, e.g.][]{1985ibs..book.....P}. However, even if the mass remains
in the system, orbital angular momentum may not be conserved
($\dot{J}_{\mathrm{orb}}\ne 0$) because of the exchange of angular momentum
between the orbit and the stellar spins via tidal torques and mass transfer
\citep{2007ApJ...655.1010G,2013A&A...557A..40D}.

In the 60s, several works from \citet{1964AcA....14..251P},
\citet{1964AcA....14..241K} and
\citet{1969Ap&SS...3..330H,1969Ap&SS...3...31H} evaluated the gravitational
force between the material leaving the inner-Lagrangian, $\mathcal{L}_{1}$,
point (the matter stream) and each star. As a particle travels, it
generates a time-varying torque on the stars, allowing for angular momentum
to be exchanged between the transferred material and the
orbit. Subsequently, \citet{2008ARep...52..680L} demonstrated that orbital
angular momentum is conserved only if the stars are point masses, if the
gravitational force between the matter stream and the stars is neglected
and if the velocity of the ejected (accreted) material is equal in
magnitude but in the opposite direction to the orbital velocity of the mass
loser (gainer; see Appendix \ref{app:Jdot_RLOF}). In reality, the velocity
of the accreted particle is determined by the initial ejection velocity,
which in turn depends on the thermal sound speed in the primary's
atmosphere and its rotation rate
\citep{1964AcA....14..231K,1975MNRAS.170..325F}.

It is widely assumed that tides enforce the synchronous rotation of the
primary with the orbit during RLOF. However, there is observational
evidence for super- and sub-synchronously rotating stars in circular
binaries
\citep[e.g.][]{1989A&A...211...56H,1990A&A...228..365A,2006ApJ...653..621M,2007A&A...467..647Y}. Furthermore,
\citet{1976ApJ...204L..29P} and \citet{1978A&A....62..317S} argued that if
the ejected material removes angular momentum faster than tides can act to
synchronize the rotation, the primary star will rotate sub-synchronously
and its Roche lobe radius will be affected
\citep[e.g.][]{1963ApJ...138.1112L,2007ApJ...660.1624S}.

A sub-synchronously rotating primary may cause the ejected material to be
accreted back onto the primary (henceforth termed `self-accretion'), which
causes the orbit to shrink even when $q<1$
\citep{2010ApJ...724..546S}. Super-synchronous rotation, on the other hand,
may have the opposite effect
\citep{1964AcA....14..241K,1964AcA....14..251P,1967SvA....11..191P}.




In this investigation, we apply the scheme of
\citet{1969Ap&SS...3..330H,1969Ap&SS...3...31H}, who derived the equations
of motion of a binary system using the theory of osculating orbital
elements. His scheme accounts for the transfer of linear momentum between
the stars, and perturbations to the orbit due to the gravitational
attraction between the stars and the mass transfer stream. Using our binary
stellar evolution code BINSTAR, we calculate the resulting evolution of
Algol binaries for a range of initial periods and masses. Currently, we
assume that none of the exchanged mass leaves the system. Torques applied
onto each star by tides and mass transfer are also included.


The paper is organized as follows. In Sect.~\ref{sec:method} we introduce
the formalism of \citet{1969Ap&SS...3..330H,1969Ap&SS...3...31H}. Our
results are presented in Sect.~\ref{sec:results}, and discussed in
Sect.~\ref{sec:discussion}. We summarise and conclude our investigations in
Sect.~\ref{sec:conclusions}.

%

\section{Computational Method} \label{sec:method}

BINSTAR is an extension of the single-star evolution code STAREVOL. Details
on the stellar input physics can be found in \citet{2010A&A...512A..10S},
and references therein, while the binary input physics is described in
\citet{2013A&A...550A.100S} and \citet{2013A&A...557A..40D}. In this
section we present our new implementation of the osculating scheme.


\subsection{Variation of the orbital parameters}

\begin{figure}
    \begin{center}
      \includegraphics[trim = 0mm 0mm 0mm
      0mm,clip,scale=0.55]{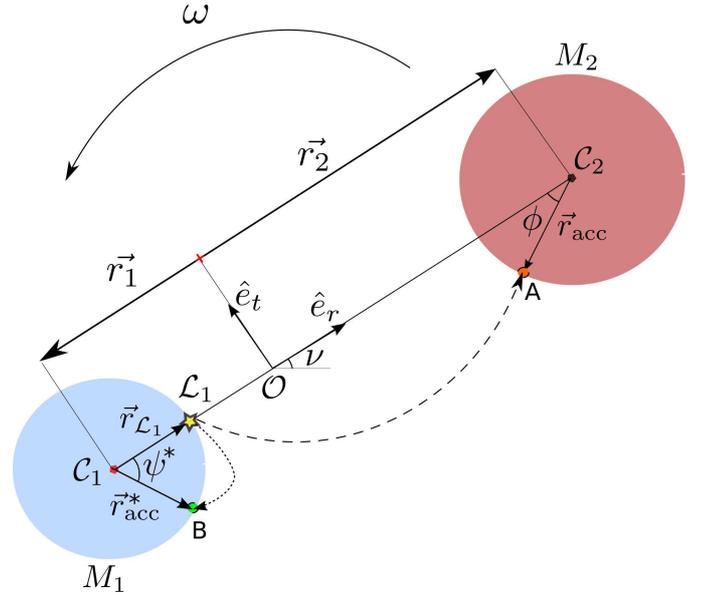}
      \caption{Schematic of a binary system, consisting of a primary star
        of mass $M_{1}$, and a secondary of mass $M_{2}$ orbiting with an
        angular speed $\omega$, where the centre of mass of the binary
        system is located at $\mathcal{O}$. The primary and secondary are
        respectively located at $\vec{r}_{1}$ and $\vec{r}_{2}$ with
        respect to $\mathcal{O}$. Mass is ejected from the
        inner-Lagrangian, $\mathcal{L}_{1}$, point (yellow star), located
        at $\vec{r}_{\mathcal{L}_{1}}$ with respect to the primary's mass
        centre, $\mathcal{C}_{1}$. Material falls towards the secondary
        (dashed line) and is accreted onto its surface (or at the edge of
        an accretion disc) at $A$, situated at $\vec{r}_{\mathrm{acc}}$
        with respect to its mass centre, $\mathcal{C}_{2}$. Alternatively,
        material falls onto the primary's surface (dotted line), landing at
        $B$, located at $\vec{r}_{\mathrm{acc}}^{*}$ with respect to
        $\mathcal{C}_{1}$. The unit vectors $\hat{e}_{r}$ and $\hat{e}_{t}$
        point along the line joining the two stars (towards the secondary),
        and perpendicular to this line, respectively, and $\nu$ is the true
        anomaly.}
      \label{fig:binary} 
    \end{center}
\end{figure}

Consider a binary system with an eccentricity $e$. The stars orbit about
their common centre of mass $\mathcal{O}$ (see Fig.~\ref{fig:binary}) with
an orbital angular speed, $\omega$, and orbital period $P_{\mathrm{orb}}$.

Material is ejected from the primary star at a rate $\dot{M}_{1}$ with a
velocity $\vec{V}_{1}$ relative to the primary's mass centre, given by
\begin{equation}
\vec{V}_{1}=\vec{W}_{1}-\vec{v}_{1},
\label{V_rel1}
\end{equation}
where $\vec{W}_{1}$ is the absolute velocity of the ejected material, and
$\vec{v}_{1}$ is the orbital velocity of the primary's mass
centre. Similarly, the velocity of the accreted material $\vec{V}_{2}$ with
respect to the secondary's mass centre is
\begin{equation}
\vec{V}_{2}=\vec{W}_{2}-\vec{v}_{2},
\label{V_rel2}
\end{equation}
where $\vec{W}_{2}$ is the absolute velocity of the accreted material, and
$\vec{v}_{2}$ is the secondary's orbital velocity. Ejection occurs from the
$\mathcal{L}_{1}$ point, located at a distance $r_{\mathcal{L}_{1}}$ from
the primary's centre. The Roche radius $R_{\mathcal{L}_{1}}$ and
$r_{\mathcal{L}_{1}}$, are calculated as in \citet{2013A&A...556A...4D},
using the formalism described by \citet{2007ApJ...660.1624S}, that accounts
for the donor's rotation.

The impact site is located at $\vec{r}_{\mathrm{acc}}$ with respect to the
secondary's centre of mass. The distance $r_{\mathrm{acc}}$ is either the
secondary's radius, $R_{2}$, for direct impact accretion or the accretion
disc radius. The latter is estimated from the distance of closest approach
of the accretion stream to the secondary, $r_{\mathrm{min}}$, and is given
by
\begin{equation}
  r_{\mathrm{acc}}=1.7r_{\mathrm{min}}
  \label{eq:rdisc}
\end{equation}
\citep{1975ApJ...198..383L,1976ApJ...206..509U}, where $r_{\mathrm{min}}$
is determined from our ballistic calculations (see Section
\ref{sub:velocities}). The vector $\vec{r}_{\mathrm{acc}}$ forms an angle
\begin{equation}
\psi=\pi-\phi,
\label{psi}
\end{equation} 
with the line joining the two stars, $\hat{e}_{r}$ (Fig.~\ref{fig:binary}).

In the theory of osculating elements, the rate of change of the semi-major
axis, $\dot{a}$, and of the eccentricity, $\dot{e}$, can be expressed as
\citep[see, e.g.][]{1960aitc.book.....S}

\begin{equation}
\dot{a}=\frac{P_{\mathrm{orb}}}{\pi(1-e^{2})^{\frac{1}{2}}}[\mathcal{S}e\sin\nu+\mathcal{T}(1+e\cos\nu)]\label{eq:adot_osc}
\end{equation}
and

\begin{equation}
\dot{e}=\frac{P_{\mathrm{orb}}(1-e^{2})^{\frac{1}{2}}}{2\pi a}\left\{
  \mathcal{S}\sin\nu+\mathcal{T}\left[\frac{2\cos\nu+e(1+\cos^{2}\nu)}{1+e\cos\nu}\right]\right\}, \label{eq:edot_osc}
\end{equation}
where $\mathcal{S}$ and $\mathcal{T}$ are the perturbing forces per unit
mass, acting along $\hat{e}_{r},$ and perpendicular to that line (along
$\hat{e}_{t}$), respectively, and $\nu$ is the true anomaly.

To account for the fact that the primary's sub-synchronous rotation may
cause a fraction $\alpha_{\mathrm{self}}$ of ejected matter to be
self-accreted, we decompose $\mathcal{S}$ and $\mathcal{T}$ into two
contributions; that resulting from accretion onto the secondary, and that
arising from self-accretion. If $\dot{M}_{1,\mathrm{ej}}^{\mathrm{self}}$
and $\dot{M}_{1,\mathrm{ej}}^{\mathrm{comp}}$ are the mass transfer rates
onto the primary and towards the secondary, respectively, then
\begin{equation}
\dot{M}_{1,\mathrm{ej}}=\dot{M}_{1,\mathrm{ej}}^{\mathrm{self}}+\dot{M}_{1,\mathrm{ej}}^{\mathrm{comp}}=\alpha_{\mathrm{self}}\dot{M}_{1,\mathrm{ej}}+(1-\alpha_{\mathrm{self}})\dot{M}_{1,\mathrm{ej}}.
\label{Mdot_tot}
\end{equation}
The mass accretion rate onto the secondary is
$\dot{M}_{2,\mathrm{acc}}=-\beta\dot{M}_{1,\mathrm{ej}}^{\mathrm{comp}}$,
where $\beta$ is the accretion efficiency ($\beta=1$ for conservative mass
transfer). The corresponding perturbing forces,
$\mathcal{S}_{\mathrm{comp}}$ and $\mathcal{T}_{\mathrm{comp}}$, are
respectively
\begin{eqnarray}
  \mathcal{S}_{\mathrm{comp}} & =
  &\left(\frac{f_{2,r}}{M_{2}}-\frac{f_{1,r}}{M_{1}}\right)_{\mathrm{comp}}+\frac{\ddot{M}_{2,\mathrm{acc}}}{M_{2}}r_{\mathrm{acc}}\cos\psi-\frac{\ddot{M}_{1,\mathrm{ej}}^{\mathrm{comp}}}{M_{1}}r_{\mathcal{L}_{1}}\nonumber
  \\ & & +\frac{\dot{M}_{2,\mathrm{acc}}}{M_{2}}(V_{2,r}-\omega
  r_{\mathrm{acc}}\sin\psi)-\frac{\dot{M}_{1,\mathrm{ej}}^{\mathrm{comp}}}{M_{1}}V_{1,r},
\label{eq:Sx}
\end{eqnarray}
and

\begin{eqnarray}
  \mathcal{T}_{\mathrm{comp}} & =
  &\underbrace{\left(\frac{f_{2,t}}{M_{2}}-\frac{f_{1,t}}{M_{1}}\right)_{\mathrm{comp}}}_{\mathcal{F}_{\mathrm{stream}}^{\mathrm{comp}}}+\underbrace{\frac{\ddot{M}_{2,\mathrm{acc}}}{M_{2}}r_{\mathrm{acc}}\sin\psi}_{\mathcal{H}_{\mathrm{comp}}}
  \nonumber \\ & & +\underbrace{\frac{\dot{M}_{2,\mathrm{acc}}}{M_{2}}(V_{2,t}+\omega
  r_{\mathrm{acc}}\cos\psi)-\frac{\dot{M}_{1,\mathrm{ej}}^{\mathrm{comp}}}{M_{1}}(V_{1,t}+\omega
  r_{\mathrm{\mathcal{L}_{1}}})}_{\mathcal{G}_{\mathrm{comp}}},
\label{eq:Ty}
\end{eqnarray}
\citep{1969Ap&SS...3...31H,2007ApJ...667.1170S}. The subscripts `$r$' and
`$t$' refer to the radial and transverse components (i.e. along
  $\hat{e}_{r}$ and $\hat{e}_{t}$) of the vector quantities,
respectively. In Eqs.~(\ref{eq:Sx}) and (\ref{eq:Ty}), terms proportional
to $\vec{f}_{i}$ are the gravitational forces per unit mass acting on the
$i$th star due to the mass transfer stream (see Section \ref{sub:stream}),
and terms proportional to $\dot{M}_{i}$ are associated with the change of
linear momentum for the $i$th star, while terms proportional to
$\ddot{M}_{i}$ represent the acceleration of the mass centre of the
$i$th star arising from asymmetric mass loss or gain. In Eq. (\ref{eq:Ty}),
$\mathcal{F}_{\mathrm{stream}}^{\mathrm{comp}}$,
$\mathcal{G}_{\mathrm{comp}}$ and $\mathcal{H}_{\mathrm{comp}}$ correspond
respectively to the gravitational force acting on the secondary by the mass
transfer stream, the linear momentum transferred to the secondary and the
acceleration of its mass centre, all with respect to the primary (see
Sect. \ref{sec:results}).

The self-accretion rate back onto the primary is
$\dot{M}_{1,\mathrm{acc}}^{\mathrm{self}}=-\dot{M}_{1,\mathrm{ej}}^{\mathrm{self}}$. The
associated perturbing forces, $\mathcal{S}_{\mathrm{self}}$ and
$\mathcal{T}_{\mathrm{self}}$, are found by considering the total
acceleration experienced by the primary resulting from the ejection and
re-capture of material (see Appendix \ref{self_accretion}), giving
\begin{eqnarray}
\mathcal{S}_{\mathrm{self}}& = &
\left(\frac{f_{2,r}}{M_{2}}-\frac{f_{1,r}}{M_{1}}\right)_{\mathrm{self}}-\frac{\ddot{M}_{1,\mathrm{ej}}^{\mathrm{self}}}{M_{1}}(r_{\mathcal{L}_{1}}-r_{\mathrm{acc}}^{*}\cos\psi^{*})\nonumber
\\ & & -\frac{\dot{M}_{1,\mathrm{ej}}^{\mathrm{self}}}{M_{1}}(V_{1,r}-V_{1,r}^{*}+\omega
r^{*}_{\mathrm{acc}}\sin\psi^{*}),
\label{eq:Sx_self}
\end{eqnarray}
and 
\begin{eqnarray}
\mathcal{T}_{\mathrm{self}}& = &
\underbrace{\left(\frac{f_{2,t}}{M_{2}}-\frac{f_{1,t}}{M_{1}}\right)_{\mathrm{self}}}_{\mathcal{F}_{\mathrm{stream}}^{\mathrm{self}}}
\underbrace{-\frac{\ddot{M}_{1,\mathrm{ej}}^{\mathrm{self}}}{M_{1}}r_{\mathrm{acc}}^{*}\sin\psi^{*}}_{\mathcal{H}_{\mathrm{self}}}\nonumber
\\ & & \underbrace{-\frac{\dot{M}_{1,\mathrm{ej}}^{\mathrm{self}}}{M_{1}}(V_{1,t}-V_{1,t}^{*}+\omega
r_{\mathcal{L}_{1}}-\omega r^{*}_{\mathrm{acc}}\cos\psi^{*})}_{\mathcal{G}_{\mathrm{self}}},
\label{eq:Ty_self}
\end{eqnarray}
where the asterisks indicate quantities calculated at self-accretion. Here,
$\mathcal{F}_{\mathrm{stream}}^{\mathrm{self}}$ is similar to
$\mathcal{F}_{\mathrm{stream}}^{\mathrm{comp}}$ but now for the
self-accreted material, while $\mathcal{G}_{\mathrm{self}}$ and
$\mathcal{H}_{\mathrm{self}}$ correspond respectively to the net momentum
transferred to the primary by the ejected and self-accreted material, and
the acceleration of its centre of mass. The radius
  $r^{*}_{\mathrm{acc}}$ is determined from the ballistic calculations, and
  corresponds to the location of the particle where the Roche potential is
  equal to the potential at the $\mathcal{L}_{1}$-point. Using
Eqs.~(\ref{eq:Sx}) to (\ref{eq:Ty_self}), the total perturbing forces are
$\mathcal{S}=\mathcal{S}_{\mathrm{comp}}+\mathcal{S}_{\mathrm{self}}$ and
$\mathcal{T}=\mathcal{T}_{\mathrm{comp}}+\mathcal{T}_{\mathrm{self}}$.

For circular orbits, $\dot{a}$ is only a function of $\mathcal{T}$. For
infinitesimal changes of $a$ and $e$ over one orbital period, the term in
braces in Eq. (\ref{eq:edot_osc}) averages to zero over one orbit, so
$\dot{e}=0$ \citep{1969Ap&SS...3...31H}. Therefore, in the remainder of
Sect. \ref{sec:method}, we will just describe the quantities pertinent to
the calculation of $\mathcal{T}$.

The total angular momentum of the binary system, $J$, is the sum of the
spin angular momenta of each star, $J_{1,2}$, the orbital angular momentum,
$J_{\mathrm{orb}}$, and the angular momentum carried by the mass that is
not attached to the stars (i.e. the mass in the wind and in the mass
transfer stream), $J_{\mathrm{MT}}$, i.e.
\begin{equation}
J=J_{1}+J_{2}+J_{\mathrm{orb}}+J_{\mathrm{MT}}=\mathrm{const}.
\label{J_sigma}
\end{equation}
The rate of change of the orbital angular momentum,
$\dot{J}_{\mathrm{orb}}$, is then determined by taking the time derivative
of Eq.~(\ref{J_sigma}), and solving for $\dot{J}_{\mathrm{orb}}$, giving
\begin{equation}
  \dot{J}_{\mathrm{orb}}=-\dot{J}_{1}-\dot{J}_{2}-\dot{J}_{\mathrm{MT}},
\label{Jdot_orb_2}
\end{equation}
where
\begin{eqnarray}
  \dot{J}_{\mathrm{MT}} & = &
  \underbrace{-J_{\mathrm{orb}}\left[\frac{\dot{M}_{1,\mathrm{ej}}^{\mathrm{comp}}}{M}\left(\frac{1}{q}-\beta{q}\right)\right]-m
    a\mathcal{T}}_{\dot{J}_{\mathrm{RLOF}}}\nonumber \\ & &
  \underbrace{-J_{\mathrm{orb}}\left(\frac{\dot{M}_{1,\mathrm{loss}}}{M}\frac{1}{q}+\frac{\dot{M}_{2,\mathrm{loss}}}{M}q \right)}_{\dot{J}_{\mathrm{lost}}}
\label{Jdot_stream}
\end{eqnarray}
for circular orbits (see Appendix \ref{app:Jdot_RLOF}). Here,
$m=M_{1}M_{2}/(M_{1}+M_{2})$ is the reduced mass,
$\dot{M}_{1,2,\mathrm{loss}}<0$ is the systemic mass loss rate from each
star (either via winds and/or non-conservative evolution),
$\dot{J}_{\mathrm{RLOF}}$ is the torque resulting from the material
transferred between the stars, and $\dot{J}_{\mathrm{lost}}$ is the torque
generated by the material leaving the system (and thus associated with
$\dot{M}_{1,2,\mathrm{loss}}$). The expression for
$\dot{J}_{\mathrm{lost}}$ is identical to that of
\citet{2008A&A...480..797B}, which considers that the escaping material
carries the specific orbital angular momentum of the star.

For brevity, we term our formalism the osculating scheme. If the stars are
point masses, and the gravitational attraction exerted by the accretion
stream is neglected, as is usually assumed, then for conservative mass
transfer $\dot{J}_{\mathrm{MT}}=0$ in Eq.~(\ref{Jdot_stream}) and we
recover the classical formulation\footnote{For conservative mass transfer
  and neglecting the stellar spins, inserting Eqs.~(\ref{Jdot_stream}) and
  (\ref{eq:edot_osc}) into Eq.~(\ref{adot_a_canonical}) recovers
  Eq.~(\ref{eq:adot_osc}) .}.


\subsection{Stellar Torques}

The torques acting on the $i$th star can be decomposed into the tidal
torque $\dot{J}_{\mathrm{tide},i}$, and the torque arising from mass
ejection or accretion, $\dot{J}_{\dot{M}_i}$, to give
\begin{equation}
\dot{J}_{i}=\dot{J}_{\mathrm{tide},i}+\dot{J}_{\dot{M}_i}.
\label{Jdot_i}
\end{equation}
For stars with radiative envelopes, we apply the prescription for dynamical
tides described by \citet{1989A&A...220..112Z}. For convective stars, we
use the formalism of \citet{1977A&A....57..383Z} describing equilibrium
tides \citep[see][for further details]{2013A&A...550A.100S}. For
$\dot{J}_{\dot{M}_{i}}$ we have
\begin{equation}
  \dot{J}_{\dot{M}_{i}}=\dot{M}_{i}(\omega r_{i}^{2}+U_{i,t}\,r_{i}\cos\phi-U_{i,r}\,r_{i}\sin\phi),
\label{J_dm}
\end{equation}
\citep{1964AcA....14..251P,1975MNRAS.170..325F} where $U_{i,t}$ and
$U_{i,r}$ are the tangential and radial components of the ejection or
accretion velocity with respect to a frame of reference co-rotating with
the binary, $r_{i}$ is the distance from the $i$th star's mass centre to
the mass ejection/accretion point (i.e. $r_{\mathcal{L}_{1}}$ or
$r_{\mathrm{acc}}$), $\phi$ is the angle between the ejection/accretion
point and the line joining the two stars, and $\dot{M}_{i}$ is the mass
loss/accretion rate. From Eq.~(\ref{J_dm}), the torque applied onto the
primary because of mass ejection is
\begin{equation}
\dot{J}_{\dot{M}_{1}}=-|\dot{M}_{1,\mathrm{ej}}|(\omega
r_{\mathcal{L}_{1}}^{2}+U_{1,t}\,r_{\mathcal{L}_{1}}),
\label{Jdot_dm1}
\end{equation}
where we have used the fact that $\phi=0$. The torque arising from
self-accretion is
\begin{equation}
\dot{J}_{\dot{M}_{1}}^{\mathrm{self}}=\dot{M}_{1,\mathrm{acc}}^{\mathrm{self}}[\omega
(r_{\mathrm{acc}}^{*})^{2}+U^{*}_{1,t}\,r^{*}_{\mathrm{acc}}\cos\phi^{*}-U^{*}_{1,r}\,r^{*}_{\mathrm{acc}}\sin\phi^{*}],
\label{Jdot_self}
\end{equation}
while the torque applied onto the secondary is
\begin{equation}
\dot{J}_{\dot{M}_{2}}=\dot{M}_{2,\mathrm{acc}}(\omega
r_{\mathrm{acc}}^{2}+U_{2,t}\,r_{\mathrm{acc}}\cos\phi-U_{2,r}\,r_{\mathrm{acc}}\sin\phi).
\label{Jdot_dm2}
\end{equation}

As shown by \cite{1981A&A...102...17P}, accretion may rapidly spin up the
secondary to its critical angular velocity
$\Omega_{2}^{\mathrm{crit}}=(GM_{2}/R_{2}^{3})^{1/2}$. \citet{2013A&A...557A..40D}
argued that super-critical rotation can be avoided either by the
interaction between the secondary and an accretion disc, or magnetic braking
(wind braking and disc-locking). For simplicity, we mimic these mechanisms
by forcing the secondary's spin to remain below
$\Omega_{2}^{\mathrm{crit}}$. This translates into an effective torque on
the gainer given by
\begin{equation}
\dot{J}_{\dot{M}_{2}}^{\mathrm{eff}}=\min\,\Biggl(\frac{I_{2}\Omega_{2}^{\mathrm{crit}}-J_{2,0}}{\Delta t},\dot{J}_{\dot{M}_{2}}\Biggr)
\label{Jdot_crit}
\end{equation}
where $\Delta t$ is the evolutionary time step, which is constrained using
the nuclear burning timescales, changes in the stars' structures, the rates
of change of the orbital parameters and the mass transfer rate
\citep[see][for further details]{2013A&A...550A.100S}. Also, $I_{2}$ is the
secondary's moment of inertia and $J_{2,0}$ is the secondary's angular
momentum at the previous time step. For simplicity, we assume that each
star rotates as a solid body, since the treatment of differential rotation
is beyond the scope of this investigation (but see
Sect. \ref{sec:discussion}).

\subsection{Ejection and accretion velocities}
\label{sub:velocities}

The components of $\vec{V}_{1}$ and $\vec{V}_{2}$ along $\hat{e}_{t}$, i.e.
$V_{1,t}$ and $V_{2,t}$, respectively are given by
\citep[][]{1969Ap&SS...3...31H}

\begin{equation}
V_{1,t}=U_{1,t}+\omega r_{\mathcal{L}_{1}},\label{eq:V1y}
\end{equation}
and
\begin{equation}
V_{2,t}=U_{2,t}+\omega r_{\mathrm{acc}}\cos\psi.\label{eq:V_2y}
\end{equation}
For particles ejected from the $\mathcal{L}_{1}$ point, we set $U_{1,r}$ to
the sound speed, $c_{\mathrm{s}}$, at the primary's photosphere, and
$U_{1,t}$ is calculated from
\begin{equation}
  U_{1,t}=r_{\mathcal{L}_{1}}(\Omega_{1}-\omega),
\label{U_1t}
\end{equation}
where $\Omega_{1}$ is the spin angular velocity of the primary. We evaluate
$\psi,$ $U_{2,r}$ and $U_{2,t}$ by solving the restricted three-body
equations of motion
\citep[e.g.][]{1969Ap&SS...3..330H,1975MNRAS.170..325F}.

\begin{figure*}
  \begin{minipage}{170mm}
    \begin{center}
      \includegraphics[scale=0.70,trim=0mm 0mm 0mm
        6mm,clip]{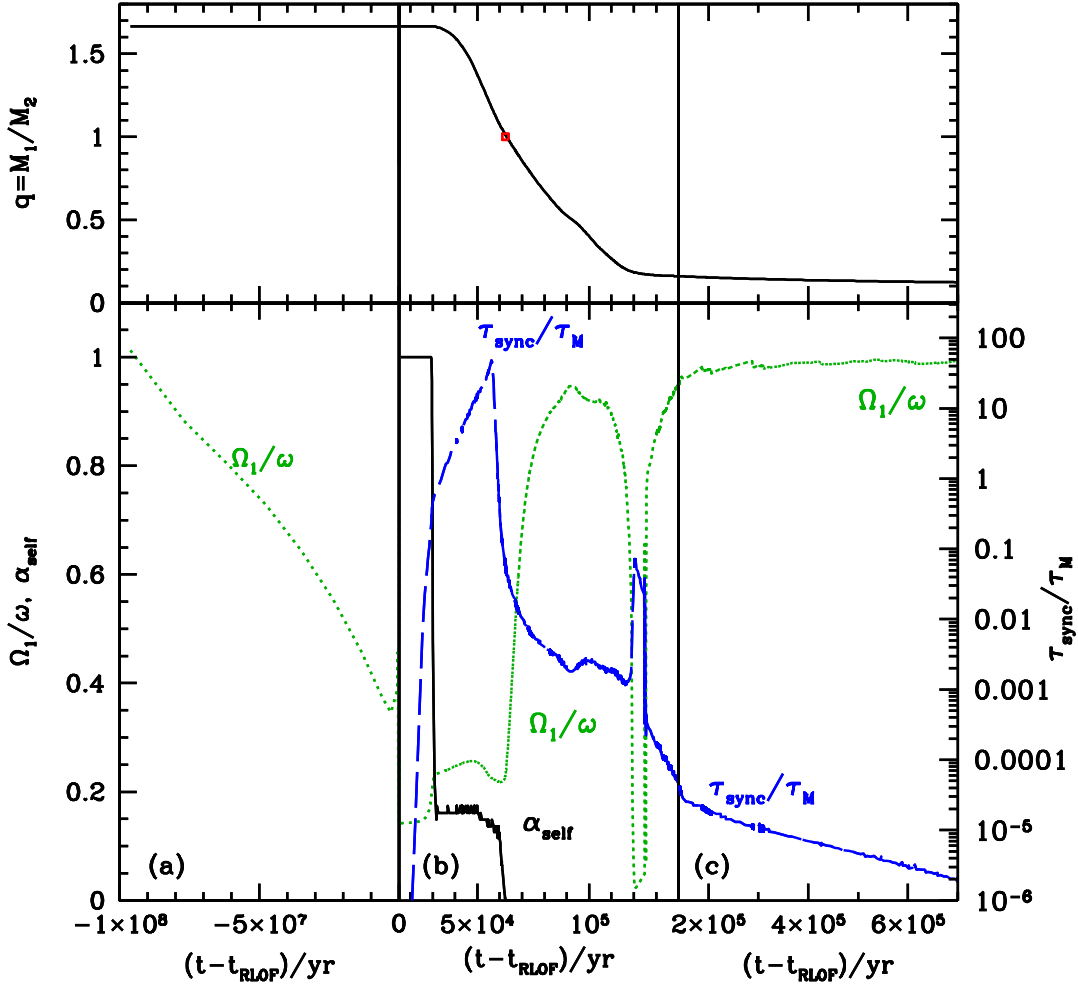}
        \caption{Evolution in time (since the start of mass transfer at
          $t_{\mathrm{RLOF}}=9.62034\times{10}^{7}$ yr) of the primary's
          spin angular velocity in units of the orbital velocity,
          $\Omega_{1}/\omega$ (dotted green curve, left axis), for the 5
          $+$ 3 $M_{\odot}$ system, initial period $P_{i}=7$ d, using
          the osculating formalism. Panel (a): as the primary evolves off
          the main sequence up to the onset of mass transfer; panel (b):
          during the self-accretion phase; panel (c): during the slow mass
          transfer phase. Panel (b) also shows the fraction of ejected
          material that falls back onto the primary,
          $\alpha_{\mathrm{self}}$ (solid black curve, left axis), and the
          ratio of the tidal synchronization timescale to the mass
          transfer timescale, $\tau_{\mathrm{sync}}/\tau_{\dot{M}}$ (dashed
          blue curve, right axis). Note the change in scales along the
          x-axis in each panel. The top panels indicate the evolution of
          $q=M_{1}/M_{2}$, and the open red square indicates where $q=1$.}
        \label{alpha_self}
    \end{center}
  \end{minipage}
\end{figure*}

\subsection{Perturbing forces from the accretion stream}
\label{sub:stream}

We discretize the mass transfer stream as a succession of individual
particles of mass $\delta m_i$. The force exerted onto the $j$th star by
the $i$th particle is

\begin{equation}
\vec{f}_{ji}=\frac{GM_{j}\,\delta m_{i}}{r_{ji}^{3}}\vec{r}_{ji},\label{eq:f_ji}
\end{equation}
where $r_{ji}$ is the distance between the $i$th particle and star
$j$. In the transverse direction, we have for the primary

\begin{equation}
r_{1i}=\left\{ [x_{i}-(1-\mu)a]^{2}+y_{i}^{2}\right\} ^{\frac{1}{2}}
\end{equation}
and for the secondary
\begin{equation}
r_{2i}=[(x_{i}-\mu a)^{2}+y_{i}^{2}]^{\frac{1}{2}}
\end{equation}
where $\mu\equiv M_{1}/(M_{1}+M_{2})$ and $(x_i,y_i)$ are the coordinates
of particle $i$ in the co-rotating frame. Summing Eq.~(\ref{eq:f_ji})
over all stream particles, and taking the difference between the
gravitational force (per unit mass) acting on the secondary and on the
primary star gives
\begin{equation}
\mathcal{F}_{\mathrm{stream}}=\frac{f_{2,t}}{M_{2}}-\frac{f_{1,t}}{M_{1}}=G\sum_{i}\delta{m}_{i}y_{i}\left(\frac{1}{r_{2i}^{3}}-\frac{1}{r_{1i}^{3}}\right).\label{eq:ft2i_ft1i}
\end{equation}
During a time-step $\delta t$ of the ballistic calculation, the mass
contained in the stream that falls towards the secondary is given by

\begin{equation}
\delta m_{i}^{\mathrm{comp}}=-\dot{M}_{1,\mathrm{ej}}^{\mathrm{comp}}\delta t.\label{eq:mi}
\end{equation}
Inserting Eq.~(\ref{eq:mi}) into Eq.~(\ref{eq:ft2i_ft1i}), then in the
limit $\delta t\rightarrow 0$,
$\mathcal{F}_{\mathrm{stream}}^{\mathrm{comp}}$ writes as
\begin{equation}
  \mathcal{F}_{\mathrm{stream}}^{\mathrm{comp}} = -G\dot{M}_{1,\mathrm{ej}}^{\mathrm{comp}}\int_{0}^{\tilde{t}_{\mathrm{comp}}}y_{\mathrm{comp}}\left(\frac{1}{r_{2}^{3}}-\frac{1}{r_{1}^{3}}\right)_{\mathrm{comp}}\,\mathrm{d}t.\label{eq:f2t_f1t_int}
\end{equation}
Similarly for the stream falling back onto the primary, we have
\begin{equation}
\delta m_{i}^{\mathrm{self}}
=-\dot{M}_{1,\mathrm{ej}}^{\mathrm{self}}\delta t \label{eq:mi_self}
\end{equation}
giving
\begin{eqnarray}
  \mathcal{F}_{\mathrm{stream}}^{\mathrm{self}} = -G\dot{M}_{1,\mathrm{ej}}^{\mathrm{self}}\int_{0}^{\tilde{t}_{\mathrm{self}}}y_{\mathrm{self}}\left(\frac{1}{r_{2}^{3}}-\frac{1}{r_{1}^{3}}\right)_{\mathrm{self}}\,\mathrm{d}t.\label{eq:f2t_f1t_int_self}
\end{eqnarray}
In Eqs. (\ref{eq:f2t_f1t_int}) and (\ref{eq:f2t_f1t_int_self}), $\tilde{t}$
is the particle's travel time between the $\mathcal{L}_{1}$ point and the
point of impact, and $r_{1}$, $r_{2}$, and $y$ describe the position of the
particle at time $t$, given by our integration of the ballistic
trajectories. The subscripts `comp' and `self' indicate quantities
pertaining to material falling onto the secondary and onto the primary,
respectively.

\subsection{Treatment of the mass transfer stream}
\label{sub:width}

The finite width of the matter stream that is ejected from the
$\mathcal{L}_{1}$-point is accounted for when calculating the quantities
found in Eqs. (\ref{eq:Ty}) and (\ref{eq:Ty_self}) as described below.  For
a primary with an effective temperature $T_{\mathrm{eff,1}}$, mean
molecular weight at the photosphere $\mu_{\mathrm{ph,1}}$, the surface
area, $\mathcal{A}$, of the stream at the $\mathcal{L}_{1}$ point is
\begin{equation}
\mathcal{A}=\frac{2\pi\mathcal{R}T_{\mathrm{eff,1}}}{\mu_{\mathrm{ph,1}}}\frac{a^{3}q}{GM_{1}}\{g(q)[g(q)-f-qf]\}^{-1/2},
\label{Q_area}
\end{equation}
\citep{2013A&A...556A...4D} where $\mathcal{R}$ is the
ideal gas constant. In Eq.~(\ref{Q_area}), 
\begin{equation}
g(q)=\frac{q}{(r_{\mathcal{L}_{1}}/a)^{3}}+\frac{1}{[1-(r_{\mathcal{L}_{1}}/a)]^{3}},
\label{g_q}
\end{equation}
\citep{1990A&A...236..385K}
and
\begin{equation}
f=\frac{\Omega_{1}}{\omega}
\label{frot}
\end{equation}
corrects for any effects arising from the primary's rotation.


We then calculate the ballistic trajectory of $N$ particles, which are
ejected at evenly spaced intervals, $\Delta\ell$, along the width of the
$\mathcal{L}_{1}$-point, $\ell\approx\sqrt{\mathcal{A}}$. To account for
the Gaussian density distribution of the particles at the $\mathcal{L}_{1}$
nozzle \citep[see,
  e.g.][]{1975ApJ...198..383L,1987MNRAS.229..383E,2012MNRAS.427.1702R} we
weight each trajectory using
\begin{equation}
  \xi(\tilde{\ell})=\eta e^{-\frac{\tilde{\ell}^{2}}{2\sigma^{2}}},
  \label{rho_dist}
\end{equation}
where $-1\le\tilde{\ell}\le 1$ is the normalised distance along the finite
width of the stream (the $\mathcal{L}_{1}$-point is located at
$\tilde{\ell}=0$), $\sigma$ is the standard deviation and $\eta$ is a
normalisation constant set by the requirement that
$\int_{-1}^{1}\xi(\tilde{\ell})\,\mathrm{d}\tilde{\ell}=1$. We use
$\sigma=0.4$ so that the density at $\tilde{\ell}=-1$ and $\tilde{\ell}=1$
is equal to the donor's photospheric density, and use $N=128$ particles. We
checked that increasing $N$ any further or slightly varying $\sigma$
has negligible impact on the calculations.

Since we are dealing with more than one particle landing on each star, we
calculate the average stream properties as follows. Let
$Q^{\mathrm{comp}}_{k}\in\{\mathcal{F}_{\mathrm{stream},k}^{\mathrm{comp}},U_{1,2,t,k},r_{\mathrm{acc},k},\psi_{k}\}$
be some quantity related to the $k$th particle that is ejected from
$\tilde{\ell}_{k}$, which is subsequently accreted by the companion.  Then
the mean value of this quantity for a given model is calculated using

\begin{equation}
\langle{Q}^{\mathrm{comp}}\rangle=\frac{\sum^{N_{\mathrm{comp}}}_{k=1}Q^{\mathrm{comp}}_{k}\xi(\tilde{\ell_{k}})\,\Delta\tilde{\ell}}{\sum^{N_{\mathrm{comp}}}_{k=1}\xi(\tilde{\ell_{k}})\,\Delta\tilde{\ell}},
\label{mean}
\end{equation}
where $N_{\mathrm{comp}}$ is the number of particles landing on the
companion. A similar expression for the particles landing on the donor is
used, where we replace `comp' with `self'. Hence, we replace
$\mathcal{F}_{\mathrm{stream}}^{\mathrm{self,comp}}$, $U_{1,2,t}$,
$U^{*}_{1,t}$, $r_{\mathrm{acc}}^{(*)}$ and $\psi^{(*)}$ in
Eqs.~(\ref{eq:Ty}) and (\ref{eq:Ty_self}) with the corresponding mean
values, as determined by Eq.~(\ref{mean}). Finally, the fraction of
particles landing back on the donor star is
\begin{equation}
\alpha_{\mathrm{self}}=\sum_{k=1}^{N_{\mathrm{self}}}\xi(\tilde{\ell_{k}})\,\Delta\tilde{\ell},
\label{alpha_self_frac}
\end{equation}
where $N_{\mathrm{self}}$ is the number of particles undergoing
self-accretion ($N=N_{\mathrm{self}}+N_{\mathrm{comp}}$).

\begin{figure*}
  \begin{minipage}{185mm}
    \begin{center}
      
      \includegraphics[scale=0.45,trim=0mm 0mm 0mm
        6mm,clip]{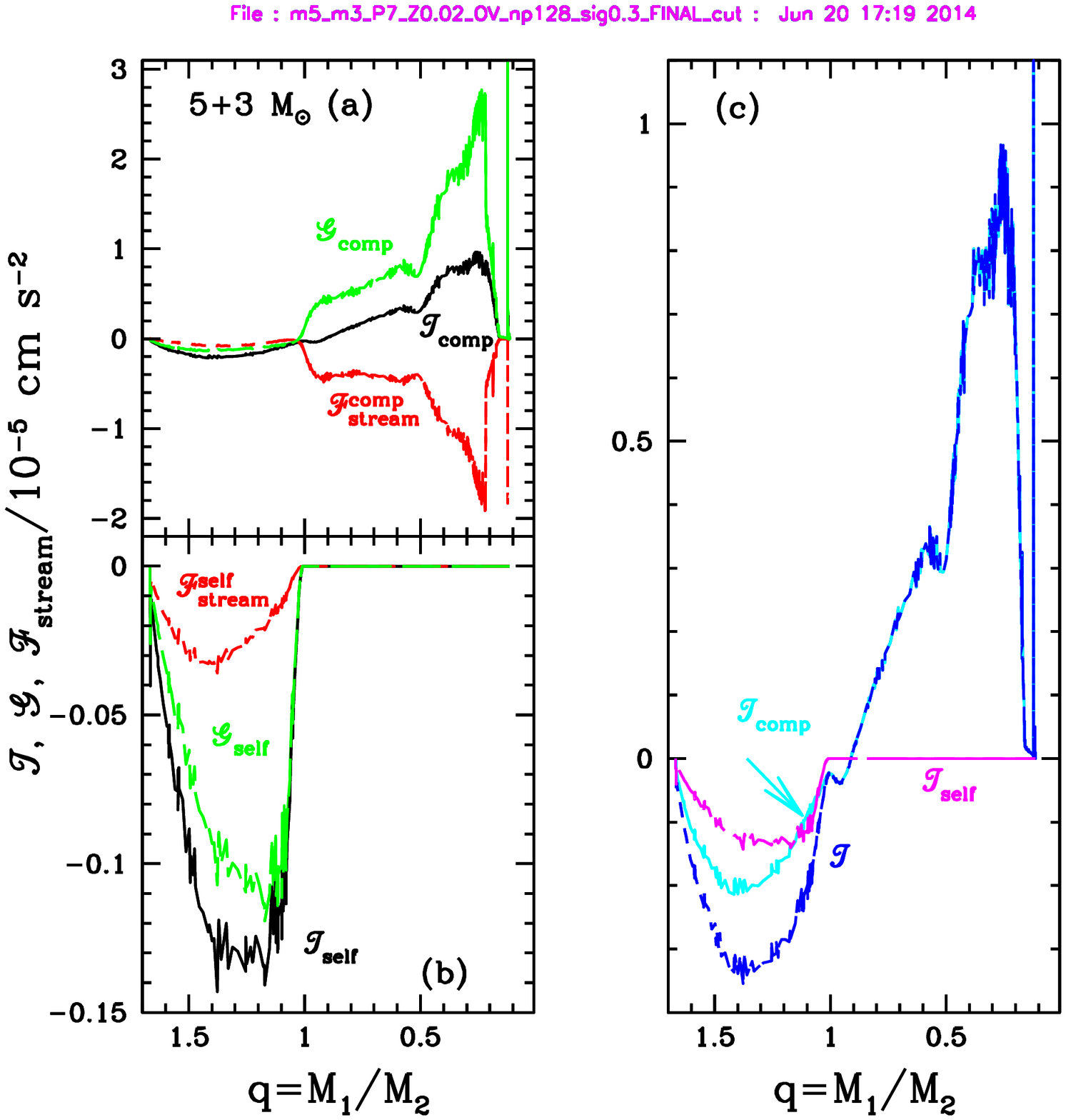}
      \includegraphics[scale=0.45,trim=0mm 0mm 0mm
        6mm,clip]{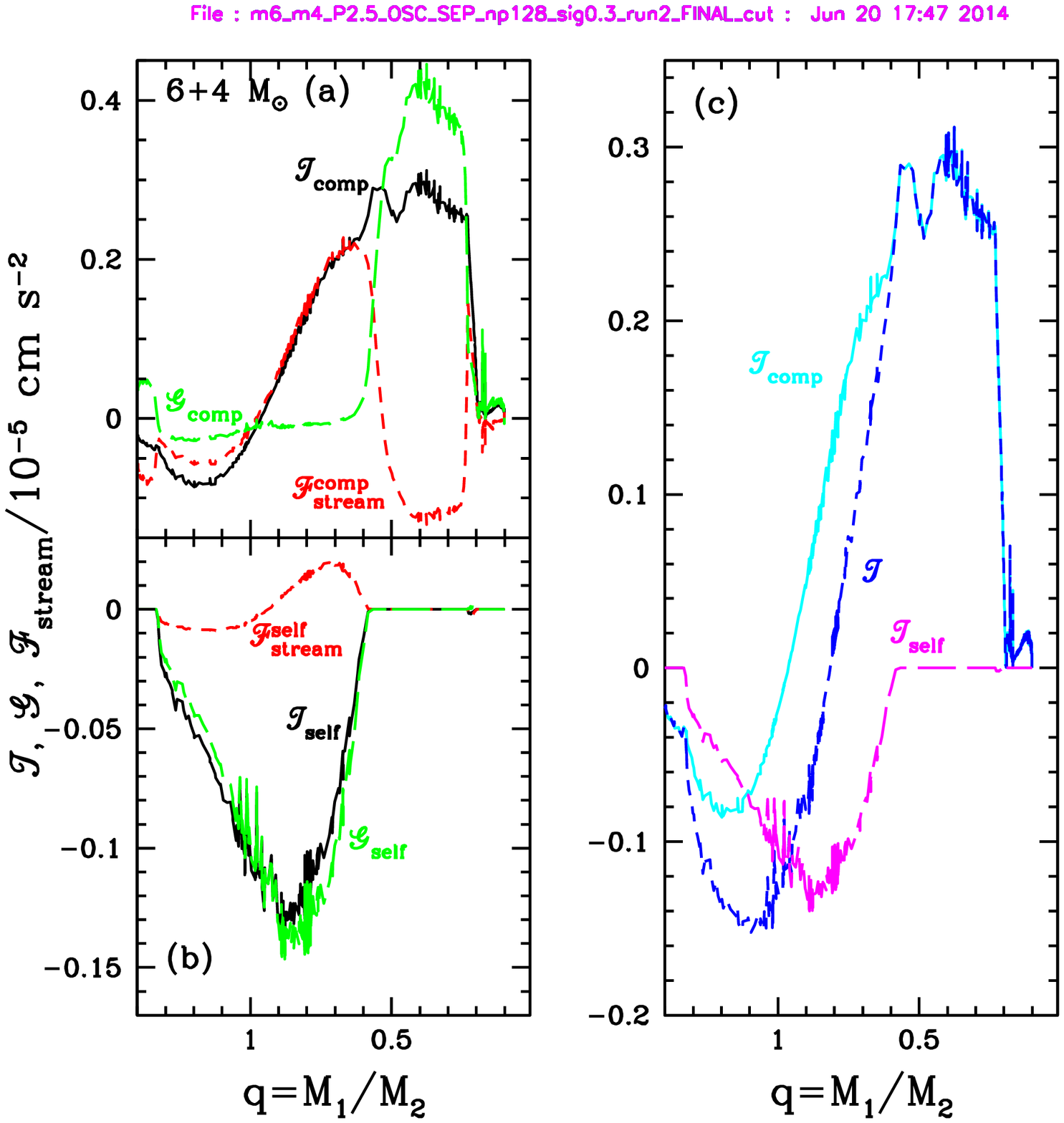}
        \caption{Left: Evolution for the 5 $+$ 3 $M_{\odot}$ binary system
          with $P_{i}=7$ d of $\mathcal{T}_{\mathrm{comp,self}}$ (solid
          black curve), $\mathcal{G}_{\mathrm{comp,self}}$ (long-dashed
          green) and $\mathcal{F}_{\mathrm{stream}}^{\mathrm{comp,self}}$
          (short-dashed red) arising from mass transfer onto the companion
          (panel (a)) and self-accretion (panel (b)). Panel (c) compares
          $\mathcal{T}_{\mathrm{self}}$ (long-dashed magenta),
          $\mathcal{T}_{\mathrm{comp}}$ (solid cyan) and the total
          $\mathcal{T}$ (short-dashed blue).  Right: the same but for the 6
          + 4 $M_{\odot}$ system, $P_{i}=2.5$ d.}
        \label{perturb_Ty}
    \end{center}
 \end{minipage}
\end{figure*}

\begin{figure*}
  \begin{minipage}{170mm}
    \begin{center}

      \includegraphics[scale=0.80,trim = 0mm 0mm 0mm 10mm,
        clip]{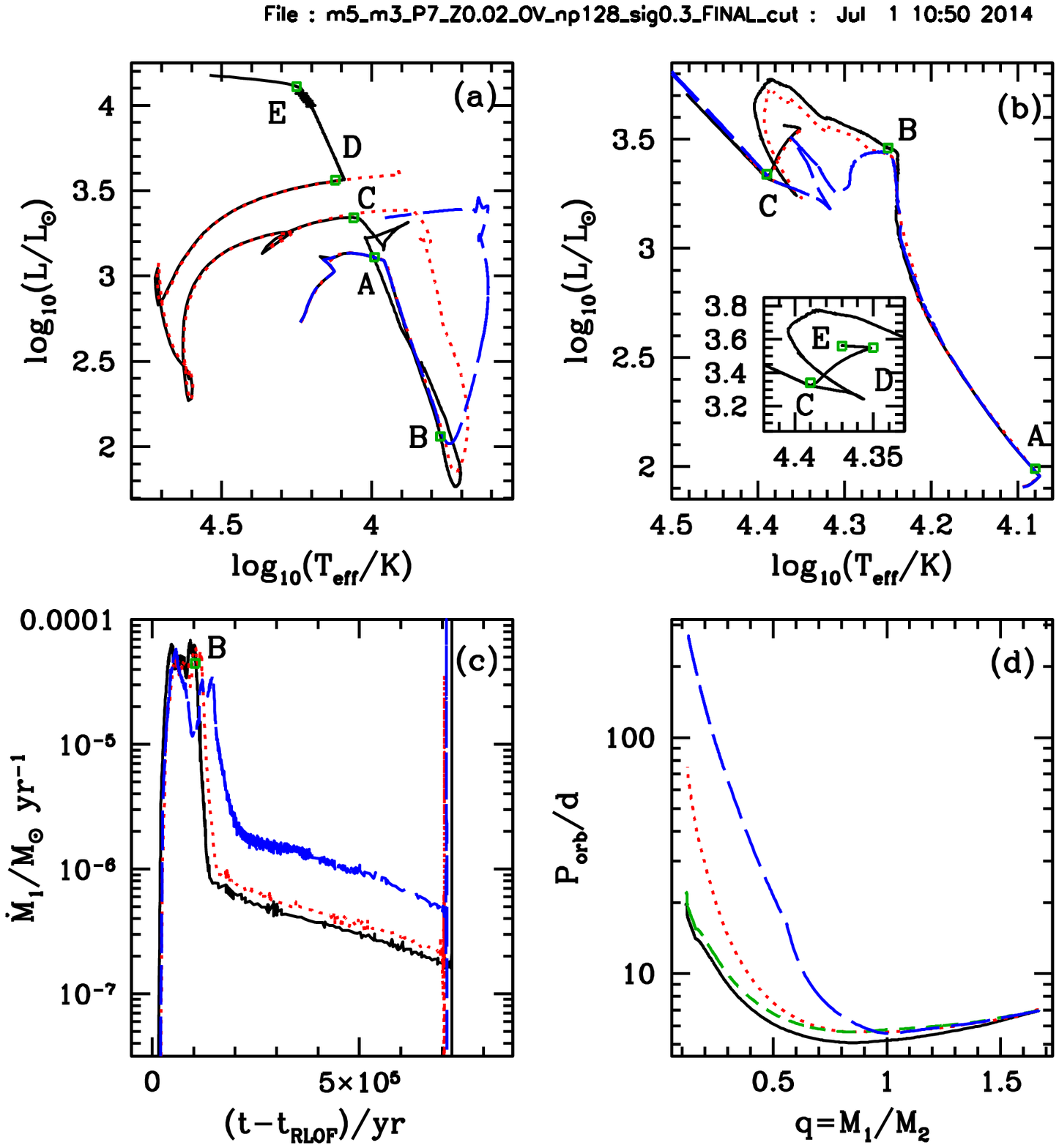}
        \caption{Evolution of the 5 $+$ 3 $M_{\odot}$ system, $P_{i}=7$ d,
          calculated using the osculating (solid black curves) and the
          classical (dotted red curves) prescriptions. The short-dashed
          green and the long-dashed blue curves also use the osculating
          scheme, but $\mathcal{T}_{\mathrm{self}}$ and
          $\mathcal{F}_{\mathrm{stream}}$ have been set to zero,
          respectively. For clarity, the dashed green curve has been
          omitted from panels (a) to (c) as it is indistinguishable from
          the black curve. The long-dashed blue curve has been truncated in
          panel (a), since it follows the same track as the black and red
          curves during core-He burning. Panels (a) and (b) show the
          evolutionary path in the Hertzsprung-Russel diagram of the
          primary and secondary. Panel (c) shows the mass loss rate,
          $\dot{M}_{1}$, as a function of time since mass transfer started
          and panel (d) the evolution of the orbital period,
          $P_{\mathrm{orb}}$, as a function of the mass ratio,
          $q=M_{1}/M_{2}$. The different evolutionary phases are indicated
          by the open green squares, and are labelled as follows: A: start
          of case B mass transfer; B: $q=1$; C: end of case B mass transfer
          (coincides with the end of H-shell burning); D: start of case BB
          mass transfer (onset of He-core burning); E: end of case BB mass
          transfer (onset of He-shell burning).}
      \label{multi}
    \end{center}
  \end{minipage}
\end{figure*}

\subsection{The binary models}

To analyse the impact of this new formalism on the evolution of Algols, we
consider two systems. The first configuration is a 5 $+$ 3 $M_{\odot}$
binary, with an initial period $P_{i}=7$ days, undergoing case B mass
transfer (during shell H-burning). Our second system is a 6 $+$ 4
$M_{\odot}$ binary, $P_{i}=2.5$ days, which commences mass exchange during
core H-burning (case A). We assume a solar composition ($Z=0.02$), and
apply moderate convective core overshooting, with
$\alpha_{\mathrm{ov}}=\Lambda/H_{\mathrm{p}}=0.2$, where $\Lambda$ is the
mixing length and $H_{\mathrm{p}}$ is the pressure scale height. The
initial spin periods of each star are set to the initial orbital
periods at the start of the simulation (i.e. $f=1$ in Eq.~(\ref{frot})) and
the orbits are circular.

%

\section{Results} \label{sec:results}

Section \ref{sec:case_B} presents our osculating calculations for the case
B system, which are compared to calculations determined from the classical
scheme (Sect. \ref{subsub:caseB}), followed by case BB (shell He-burning)
mass transfer in Sect. \ref{sec:case_BB}. Calculations for the case A model
are given in Sect. \ref{sec:case_A}.

\subsection{5+3 $M_{\odot}$, $P_{\mathrm{i}}=7$ days}
\label{sec:case_B}

\subsubsection{Case B mass transfer}
\label{subsub:caseB}

As the primary evolves off the main sequence, its radius and moment of
inertia increase on a much shorter timescale ($\tau_{R}=R/\dot{R}\approx
10^{8}$ yr) than tides can maintain synchronization
($\tau_{\mathrm{sync}}=|\Omega_{1}-\omega|/\dot{\omega}\approx 10^{10}$ yr)
and $\Omega_{1}$ declines (Fig.~\ref{alpha_self}a). Therefore, mass
transfer starts while the primary star is rotating significantly
sub-synchronously, and this has two important consequences. First, the Roche
radius calculated using the \citet{2007ApJ...660.1624S} formalism is 7 per
cent larger than the \citet{1983ApJ...268..368E}
prescription. Consequently, the primary must evolve further along the
sub-giant branch before it can start transferring mass.
Secondly, at the onset of RLOF, all the material ejected by the primary
star is initially self-accreted ($\alpha_{\mathrm{self}}=1$, panel
(b), solid black curve). This process induces a positive torque onto
the primary (Eq.~\ref{Jdot_self}), increasing $\Omega_{1}$ (dotted green
curve) and thus $U_{1,t}$, causing material to progressively flow onto the
companion ($\alpha_{\mathrm{self}}<1$). In parallel, mass ejected through
the $\mathcal{L}_{1}$-point applies a negative torque onto the primary
(Eq.~\ref{Jdot_dm1}). Since, initially, more material is falling onto the
primary than onto the companion, the net effect is an acceleration of the
primary's rotation rate,
i.e. $|\dot{J}_{\dot{M}_{1}}|<\dot{J}_{\dot{M}_{1}}^{\mathrm{self}}$, and a
reduction in the amount of self-accreted material. As $\Omega_{1}$
increases, $\dot{J}_{\dot{M}_{1}}^{\mathrm{self}}$ drops until eventually
the net torque applied onto the primary via the ejection and self-accretion
process is zero,
i.e. $|\dot{J}_{\dot{M}_{1}}|\approx\dot{J}_{\dot{M}_{1}}^{\mathrm{self}}$. This
situation is characterized by a plateau in $\alpha_{\mathrm{self}}\approx
0.2$. Throughout the self-accretion process, the mass loss timescale
$\tau_{\dot{M}}=M_{1}/|\dot{M}_{1}|<\tau_{\mathrm{sync}}$, so tides are
unable to enforce synchronous rotation (dashed, blue curve).

For circular orbits, Eq.~(\ref{eq:adot_osc}) reduces to
\begin{equation}
\dot{a}=\frac{P_{\mathrm{orb}}}{\pi}\mathcal{T}=\frac{P_{\mathrm{orb}}}{\pi}(\mathcal{T}_{\mathrm{comp}}+\mathcal{T}_{\mathrm{self}}),
\label{adot_osc_2}
\end{equation}
and so the sign of $\dot{a}$ depends on the signs of
$\mathcal{T}_{\mathrm{comp}}$ and $\mathcal{T}_{\mathrm{self}}$. To
understand the evolution of the separation, we show in
Fig.~\ref{perturb_Ty} the various contributions entering the expression for
$\cal{T}$, i.e. the gravitational force applied onto the secondary
(primary) by the ejected particle
$\mathcal{F}_{\mathrm{stream}}^{\mathrm{comp,(self)}}$, and the linear
momentum transferred to the companion (donor),
$\mathcal{G}_{\mathrm{comp,(self)}}$, all with respect to the
donor. Throughout mass transfer, the terms $\mathcal{H}_{\mathrm{self}}$
and $\mathcal{H}_{\mathrm{comp}}$, corresponding to the acceleration of
  the donor's or companion's mass centre, are negligible because
$\ddot{M}_{1,\mathrm{ej}}\approx 0$, and for this reason it is not
displayed in Fig~\ref{perturb_Ty}. At the start of mass transfer, the
contribution to $\mathcal{T}_{\mathrm{comp}}$ partially comes from
$\mathcal{F}_{\mathrm{stream}}^{\mathrm{comp}}$ (Fig. \ref{perturb_Ty}a,
left panels, red short-dashed curves), which is negative for two
reasons. Firstly, the particles are located at $y_{\mathrm{comp}}<0$ and
secondly they are typically situated in the vicinity of the secondary
i.e. $1/r_{2}>1/r_{1}$ (see Eq.~\ref{eq:f2t_f1t_int}).

The $\mathcal{G}_{\mathrm{comp}}$ term (green, long-dashed curves) is small
for $q>1$, but increases as $q$ declines. Indeed, as mass transfer
proceeds, the primary's Roche lobe radius and therefore
$r_{\mathcal{L}_{1}}$ shrink, and so the $\mathcal{L}_{1}$ point moves
further away from the companion's surface. A particle's travel time thus
rises and it can accelerate to a larger impact velocity ($V_{2,t}$). Even
though $\mathcal{G}_{\mathrm{comp}}$ also depends on $V_{1,t}+\omega
r_{\mathcal{L}_{1}}$, its magnitude is about a factor of 10 smaller than
the $|V_{2,t}+\omega r_{\mathrm{acc}} \cos\psi|$ term.

During self-accretion, $\mathcal{T}_{\mathrm{self}}$ is dictated by
$\mathcal{G}_{\mathrm{self}}$ (Fig.~\ref{perturb_Ty}b). We find that
$V_{1,t}\approx -1\times{10}^{7}$ cm s$^{-1}$ and
$V_{1,t}^{*}\approx 8\times{10}^{6}$ cm s$^{-1}$, which, for small
$\psi^{*}$, gives $\mathcal{T}_{\mathrm{self}}<0$
(Eq.~\ref{eq:Ty_self}). Similarly to
$\mathcal{F}_{\mathrm{stream}}^{\mathrm{comp}}$,
$\mathcal{F}_{\mathrm{stream}}^{\mathrm{self}}<0$ because
$y_{\mathrm{self}}<0$ and, owing to the large value of
$r_{\mathcal{L}_{1}}$ for $q>1$, the particle is closer to the
secondary than to the primary.

Hence, self-accretion enhances the rate of shrinkage of both the orbital
separation (Fig.~\ref{perturb_Ty}c) and the primary's Roche radius. This
faster contraction of $R_{\mathcal{L}_{1}}$ increases the overfilling
factor resulting in a mass transfer rate that is a factor of $\sim 1.6$
higher than in the classical scheme. In response to this higher mass loss
rate, the primary's radiative layers further contract and it attains a
lower surface luminosity on the Hertzsprung-Russel (HR) diagram compared to
the classical calculation (Fig. \ref{multi}a).

To quantify the impact of self-accretion, we re-ran a simulation by setting
$\mathcal{T}_{\mathrm{self}}=0$, as indicated by the short-dashed green
curve in Fig.~\ref{multi}d. Differences in the mass transfer rate and the
evolution along the HR diagram are negligible. However, the post-mass
transfer orbital period (about 21 d) is slightly longer than when
self-accretion is included (17 d); a relative difference of about 20 per
cent.


As mass transfer proceeds, $\tau_{\mathrm{sync}}$ decreases as a result of
the deepening of the primary's surface convection zone in response to mass
loss \citep{1977ApJ...211..486W}. Eventually,
$\tau_{\mathrm{sync}}<\tau_{\dot{M}}$ (Fig.~\ref{alpha_self}b), and tides
can re-synchronize the primary. Moreover, as the primary's mass declines,
it exerts a smaller gravitational attraction onto the ejected particle, and
evermore material falls onto the secondary star, as indicated by the
decrease in $\alpha_{\mathrm{self}}$.

Once $q\lesssim 1$, self-accretion shuts off and soon after (when $q \simeq
0.9$) $\mathcal{T}_{\mathrm{comp}}$ becomes positive, allowing the orbit to
expand. The rise in $\mathcal{T}_{\mathrm{comp}}$ is because of the
aforementioned growth of the $\mathcal{G}_{\mathrm{comp}}$ term, resulting
from the higher impact velocity ($V_{2,t}$).  Eventually, the primary
restores thermal equilibrium, $\dot{M}_{1}$ declines (Fig.~\ref{multi}c)
and the evolution enters the slow phase (around $q \approx 0.16$), where
mass transfer occurs on the nuclear timescale of the hydrogen-burning shell
\citep{1971ARA&A...9..183P}. The calculated mass transfer rates for the
classical and osculating models are virtually identical, since the
shell-burning properties in both cases are the same\footnote{The spike just
  before mass transfer terminates is due to the ignition of core Helium
  burning.}. Also note that throughout the slow phase,
$\tau_{\mathrm{sync}}\ll \tau_{\dot{M}}$ and so tides can enforce
synchronous rotation of the primary (Fig. \ref{alpha_self}c).

By the end of the simulations, the difference in the periods is significant
with 17 days for the osculating scheme compared to 71 days for the
classical model. The primary's radius is correspondingly smaller, since it
keeps track of its Roche lobe, which is a function of $a$. This explains
why, for a given luminosity, the osculating model gives a hotter primary
than in the classical case (Fig.~\ref{multi}a).

The shorter orbital period predicted by the osculating scheme is caused by
the negative $\mathcal{F}_{\mathrm{stream}}^{\mathrm{comp}}$
contribution. Indeed, neglecting this term gives
$\mathcal{T}_{\mathrm{comp}}=\mathcal{G}_{\mathrm{comp}}$, yielding the
{\it longest} post-mass transfer orbital periods out of all our
calculations (Fig.~\ref{multi}d, long-dashed blue curve). In this case, the
Roche filling primary has a larger radius and a lower effective
temperature, causing a substantial surface convection zone to develop
(radial extent of about 60 $R_{\odot}$). This enhances the mass transfer
rate because convective layers expand upon mass loss, causing the star to
over-fill its Roche lobe further.

When He ignites in the primary, mass transfer terminates and the final
masses are virtually identical between the classical and osculating schemes
($M_{1}\approx 0.87$ $M_{\odot}$, $M_{2}\approx 7.3$ $M_{\odot}$).



Both the osculating and classical formalisms predict similar evolutionary
tracks of the secondary on the HR diagram (Fig \ref{multi}b). Once mass
transfer enters the slow phase, the secondary relaxes towards thermal
equilibrium, establishing a new effective temperature and luminosity along
the main sequence which is appropriate for its new mass.


\subsubsection{Case BB mass transfer}
\label{sec:case_BB}

\begin{figure}
    \begin{center}
      \includegraphics[scale=0.57,trim=0mm 0mm 00mm
        70mm,clip]{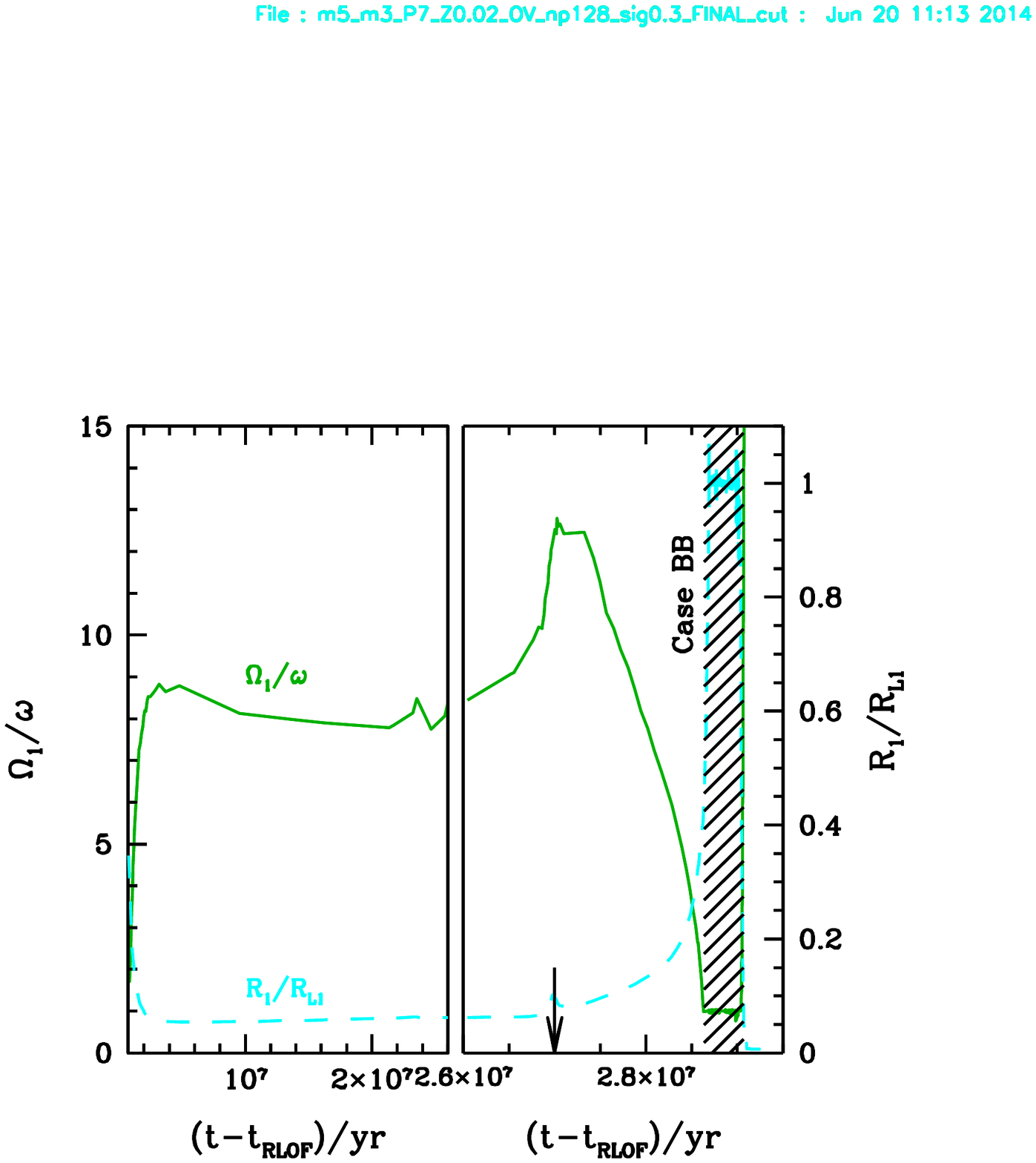}
        \caption{Similar to Fig.~\ref{alpha_self}, but now during core He
          burning (left panel), and shell He-burning (right panel), the
          onset of which is indicated by the arrow. The dashed cyan curve
          shows $R_{1}/R_{\mathcal{L}1}$ (right axis), and the shaded
          region marks case BB mass transfer.}
      \label{spins_HeB}
    \end{center}
\end{figure}

Once mass transfer has stopped, the structure of the primary consists of a
convective He-burning core of 0.27 $M_{\odot}$, surrounded by a radiative
envelope of $\sim 0.60$ $M_{\odot}$. The H-burning shell is located at mass
coordinate $M_{r}\approx 0.73 M_{\odot}$, and has a mass of about $0.06$
$M_{\odot}$. When He ignites in the core, the primary contracts within its
Roche lobe on a timescale much shorter than $\tau_{\mathrm{sync}}$ leading
to super-synchronous rotation (Fig.~\ref{spins_HeB}, left panel, solid
green curve). The star accelerates up to approximately 3 per cent of the
critical velocity when $\Omega_1/\omega \approx 12$. These calculations
therefore predict the presence of rapidly rotating core-He burning stars in
detached binaries. To the best of our knowledge, there are no available
observations of such systems in this evolutionary phase.

At about $2.72\times{10}^{7}$ yr since the start of case B mass transfer,
the activation of shell-He burning produces a rapid expansion of the
primary. As the star fills more of its Roche lobe
($R_{1}/R_{\mathcal{L},1}$, dashed cyan curve) the tidal forces strengthen
and within $\sim 1.5\times 10^6$ yr, the primary is re-synchronized. The
ensuing case BB (Fig.~\ref{spins_HeB}, shaded region) is characterized by a
constant mass exchange rate of about $10^{-7}$ $M_{\odot}$ yr$^{-1}$, and
an expansion of the orbit because $q<1$. Synchronous rotation is maintained
during the entire phase of mass transfer, so no self-accretion occurs. In
both schemes, mass transfer lasts for about $3\times{10}^{5}$ yr.

Mass transfer ceases as a result of re-ignition of the H-burning shell,
which causes the primary's radius to shrink. At this point, the binary
consists of a 0.8 $M_{\odot}$ CO star, undergoing H-shell burning in the
surface layers, and a 7.2 $M_{\odot}$ main sequence companion. The final
orbital periods are approximately 80 days and 19 days for the classical and
osculating calculations, respectively.

To calculate the subsequent evolution of this system is beyond the scope of
the investigation. Nonetheless, we can infer its fate based on the binary
parameters. Eventually, the secondary star will evolve off the main
sequence, fill its own Roche lobe and transfer mass back to the CO primary
star. For the osculating scheme, we estimate that the secondary will fill
its Roche lobe with a radius of about 34 $R_{\odot}$ as it is crossing the
sub-giant branch, and for the classical scheme, when the star approaches
the base of the giant branch with a radius of $\sim 60 R_{\odot}$. Because
of the high mass ratio $M_{2}/M_{1}\approx 8$, which lies well above the
critical value of between 1.2 and 1.3 for dynamically unstable mass
transfer \citep{2008ASSL..352..233W}, we therefore expect this system to
enter common envelope evolution.

\begin{figure*}
  \begin{minipage}{170mm}
    \begin{center}
      \includegraphics[scale=0.40,trim=0mm 0mm 0mm
        7mm,clip]{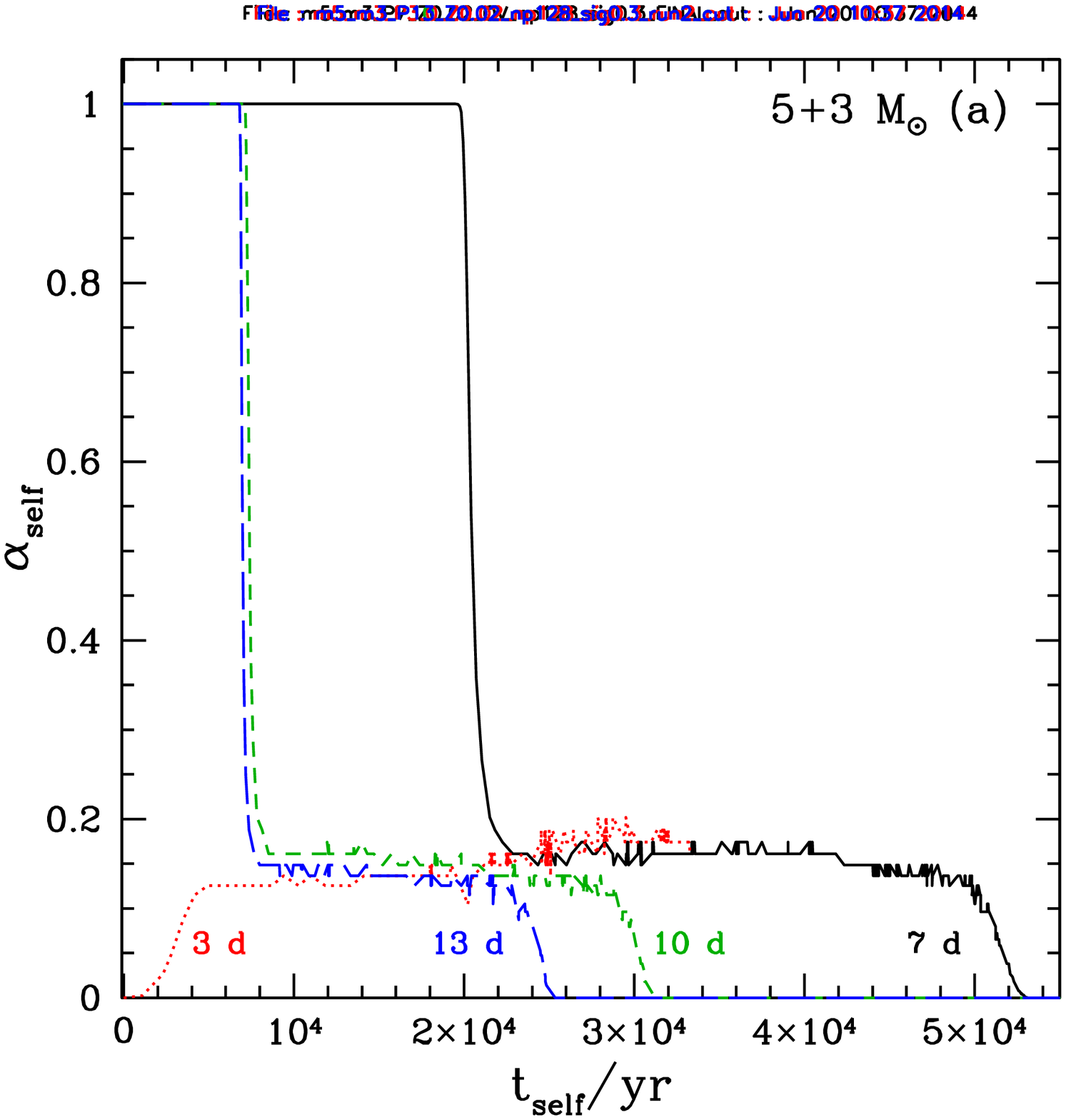}
      \includegraphics[scale=0.40,trim=0mm 0mm 0mm
        6mm,clip]{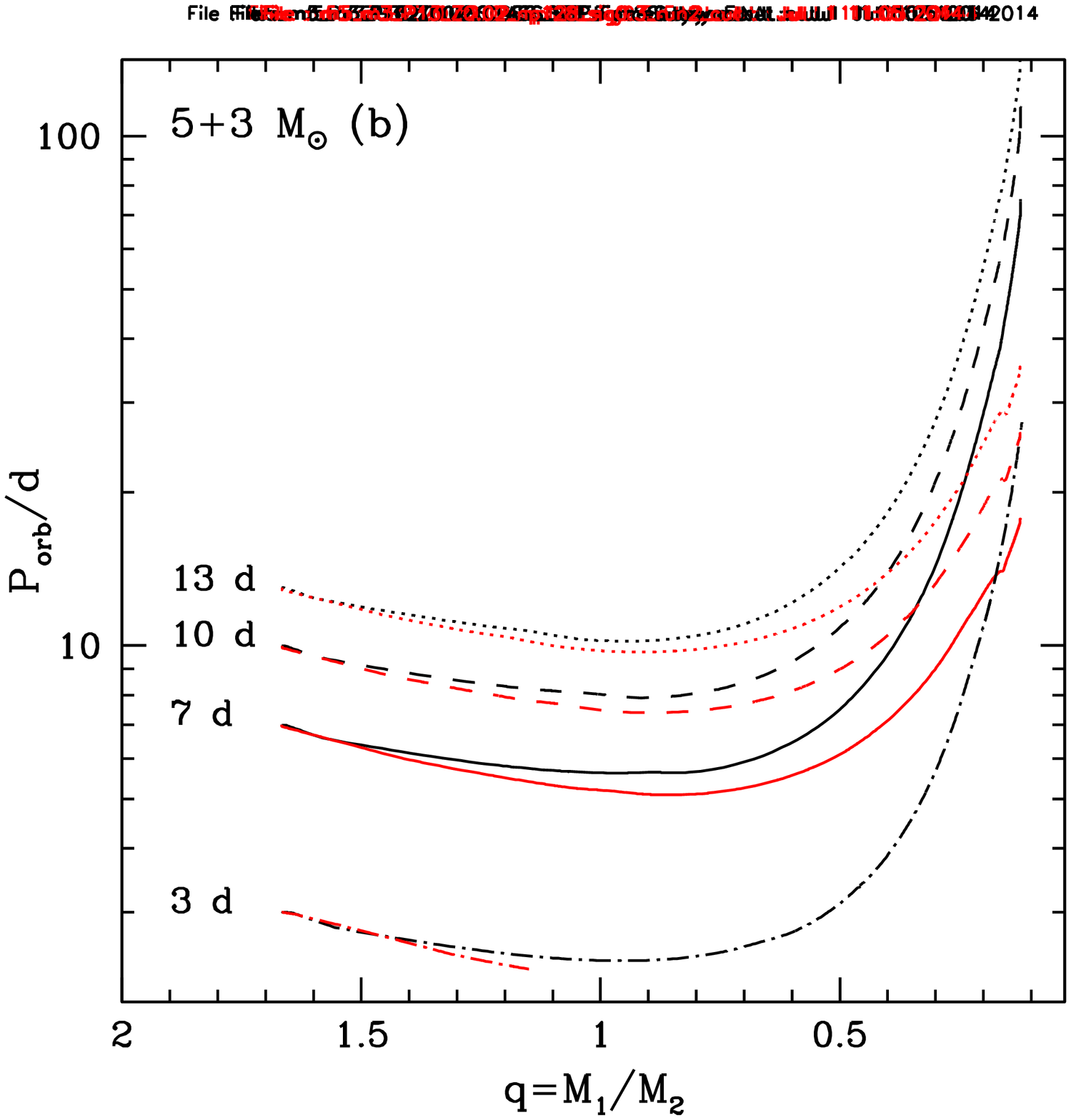}        
        \caption{(a): Evolution of $\alpha_{\mathrm{self}}$ as a function
          of time, $t_{\mathrm{self}}$, since the start of self-accretion
          for a 5 + 3 M$_{\odot}$ system, with an initial orbital
          period, $P_{i}$ of 3 d (dotted red curve), 7 d (solid black), 10
          d (green short dashed) and 13 d (blue long-dashed). (b):
          evolution of the orbital period, $P_{\mathrm{orb}}$, as a
          function of $q=M_{1}/M_{2}$, for different initial orbital
          periods: 3 d (dot-dashed); 7 d (solid); 10 d (short-dashed) and
          13 d (dotted). Black curves refer to classical calculations, and
          the red lines to the osculating scheme.}
      \label{alpha_self_Pi}
    \end{center}
 \end{minipage}
\end{figure*}

\subsubsection{Effect of changing the initial orbital period}
\label{sec:Pi}

Fig.~\ref{alpha_self_Pi}a shows that $\alpha_{\mathrm{self}}$ levels off to
between 0.15 and 0.2 for the considered values of $P_{i}$. In addition, the
duration of the self-accretion phase progressively decreases, from
$5.2\times{10}^{4}$ to $2.6\times{10}^{4}$ yr when the initial
period, $P_{i}$, rises from 7 to 13~d. By increasing $P_{i}$, the primary
evolves further along the sub-giant branch before mass transfer starts, and
can develop a deeper convection zone. When RLOF occurs, tides are more
efficient at re-synchronizing the primary, and self-accretion is stopped
earlier.

For $P_{i}$ between 7 and 13 days, $\alpha_{\mathrm{self}}\approx 1$ at the
start of mass transfer. By contrast, for $P_{i}=3$ d,
$\alpha_{\mathrm{self}}$ rises from zero to a constant value of
approximately 0.2. In this case, the primary is still very close to its
main sequence location in the HR diagram, and it is initially not rotating
sufficiently slowly to trigger self-accretion. Only once enough angular
momentum has been removed from the primary does self-accretion occur.

The final primary masses are between 0.85 and 0.88 $M_{\odot}$ and the
secondary masses between 7.15 and 7.12 $M_{\odot}$, irrespective of whether
the classical or the osculating scheme is used. However, the osculating
scheme systematically yields shorter post-mass transfer orbital periods, by
a factor of about 4 (Fig.~\ref{alpha_self_Pi}b), and for the model with
$P_{i}=3$~d, it gives rise to a contact system, in contrast to the
classical scheme.


\subsection{6+4 $M_{\odot}$, $P_{\mathrm{i}}=2.5$~d}
\label{sec:case_A}

\begin{figure*}
  \begin{minipage}{170mm}
    \begin{center}
      \includegraphics[scale=0.80,trim=0mm 0mm 0mm
        6mm,clip]{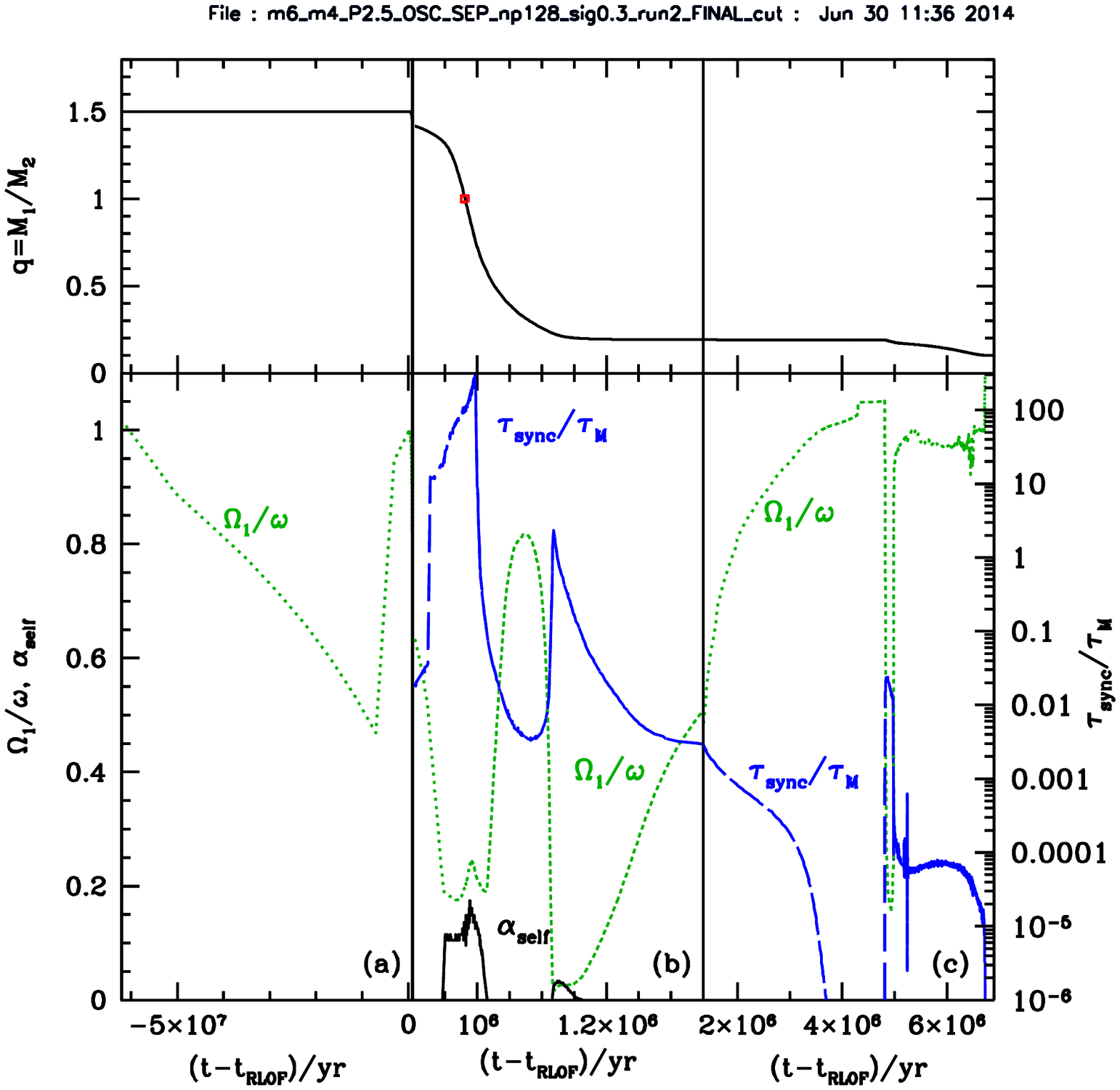}
        \caption{Similar to Fig.~\ref{alpha_self}, but now for the 6 $+$
          4 $M_{\odot}$ system, $P_{i}=2.5$ days. Here,
          $t_{\mathrm{RLOF}}=6.12305\times{10}^{7}$ yr.}
        \label{alpha_self2}
    \end{center}
  \end{minipage}
\end{figure*}

Since the initial period is shorter for this model, tides are able to
establish synchronous rotation by the time the primary fills its Roche lobe
(Fig.~\ref{alpha_self2}a). Therefore, all ejected mass is initially
accreted onto the companion ($\alpha_{\mathrm{self}}=0$) and only once mass
ejection has removed a sufficient amount of angular momentum from the
primary does self-accretion occur, with $\alpha_{\mathrm{self}}\approx
0.15$.


As we remarked, self-accretion enhances the orbital contraction because of
the negative contribution from $\mathcal{G}_{\mathrm{self}}$ (right panel,
Fig.~\ref{perturb_Ty}b). In contrast to the case B model, this process
continues after the mass ratio has reversed and only ceases when $q\approx
0.6$. The reason for this difference stems from the time delay associated
with the appearance of a surface convection zone, which reinforces the
tidal interaction and accelerates the primary's rotation velocity back to
synchronous rotation. This persisting self-accretion episode maintains
$\mathcal{T}_{\mathrm{self}}<0$ for a longer period of time and the orbit
contracts until $q\approx 0.8$ (Fig. \ref{perturb_Ty}c). Also note that
$\mathcal{F}_{\mathrm{stream}}^{\mathrm{self}}>0$ when $q<1$ because the
ejected material is typically located in the vicinity of the primary
($1/r_{1}>1/r_{2}$ in Eq.~\ref{eq:f2t_f1t_int_self}), owing to the close
proximity of the $\mathcal{L}_{1}$-point to the primary. The positive value
for $\mathcal{F}_{\mathrm{stream}}^{\mathrm{comp}}$, on the other hand, is
because the primary's sub-synchronous rotation deflects the particles such
that they have $y_{\mathrm{comp}}>0$.

\begin{figure*}
  \begin{minipage}{170mm}
    \begin{center}

      \includegraphics[scale=0.80,trim = 0mm 0mm 20mm 10mm,
        clip]{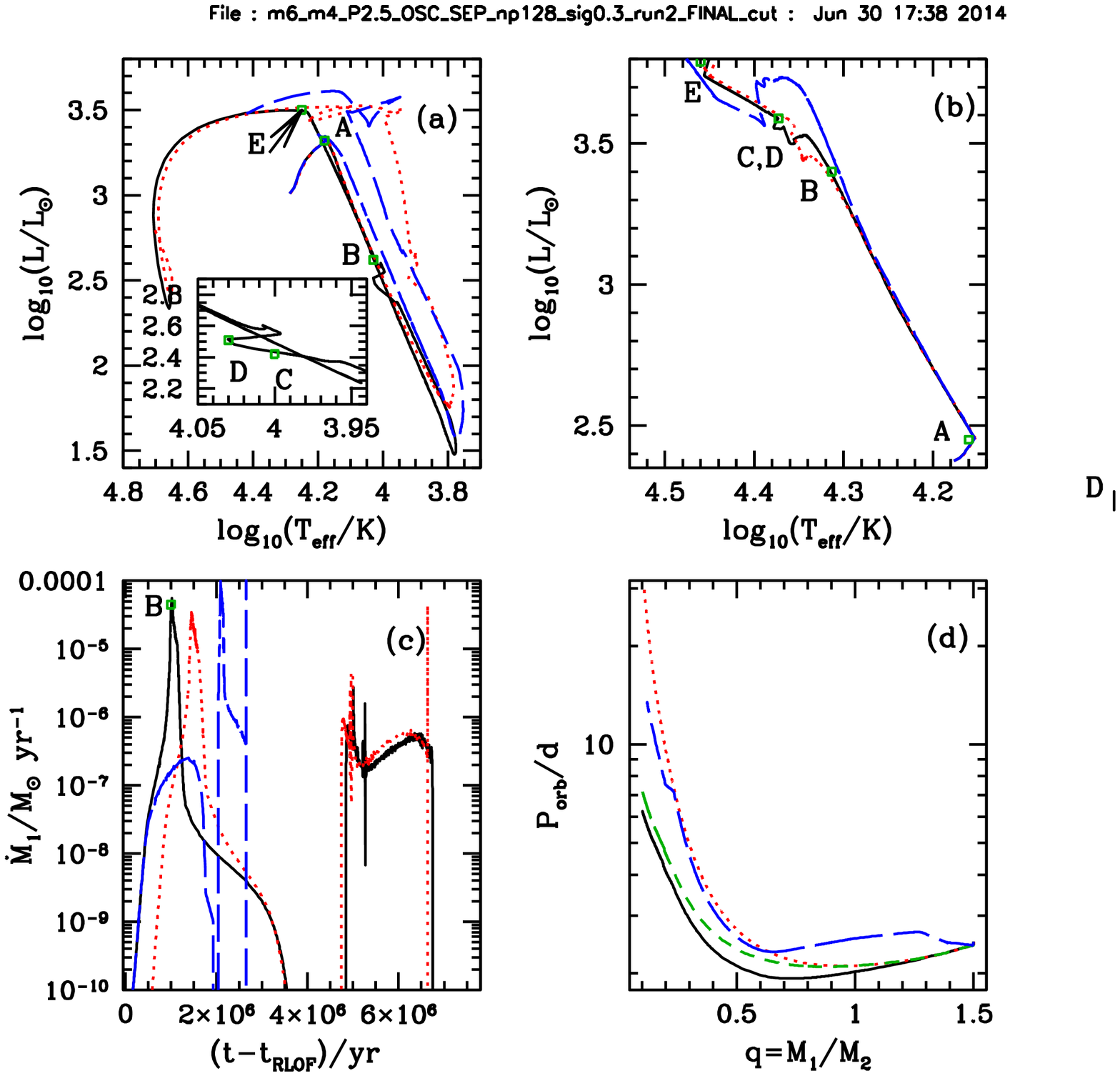}
        \caption{Similar to Fig.~\ref{multi} but for the 6 $+$ 4
          $M_{\odot}$, $P_{i}=2.5$ d system. A: start of case A mass
          transfer; B: $q=1$; C: end of core H-burning; D: start of shell
          H-burning; E: end of case A mass transfer (H-shell burning
          ceases).}
      \label{multi2}
    \end{center}
  \end{minipage}
\end{figure*}

At $t\approx t_{\mathrm{RLOF}}+1\times 10^6$ yr ($q \approx 0.3)$, the
convection zone recedes and $\tau_{\mathrm{sync}}$ rises again
(Fig.~\ref{alpha_self2}b). Subsequent mass ejection brings the primary back
into sub-synchronous rotation, triggering a second self-accretion episode,
with about 10 per cent of the ejected material falling back onto the
star. This second self-accretion event is absent in the case B model
because of the higher sound speed ($U_{1,r}$) for the primary. The impact
of this second occurrence on the orbital evolution, however, is
negligible. This is because $\dot{M}_{1,\mathrm{ej}}^{\mathrm{self}}$ and
$V_{1,t}-V^{*}_{1,t}$ are respectively a factor of about 10 and 6 smaller
than the values for the first self-accretion episode. Neglecting
$\mathcal{T}_{\mathrm{self}}$ (short-dashed green curve,
Fig.~\ref{multi2}d) shows that the relative difference between the post-mass
transfer orbital periods with and without self-accretion is about 15 per
cent.



As the mass transfer rate decelerates,
$\tau_{\mathrm{sync}}/\tau_{\dot{M}}$ correspondingly declines,
$\Omega_{1}$ increases back towards synchronous rotation and self-accretion
shuts off. Eventually, hydrogen is exhausted in the core, the primary
shrinks within its Roche lobe, and mass transfer is terminated (giving rise
to the hook feature around $\log_{10}(L/L_{\odot}) \simeq 2.5$ in
Fig.~\ref{multi2}a). Mass exchange resumes once the primary re-expands
because of H-shell ignition. Since now
$\tau_{\mathrm{sync}}/\tau_{\dot{M}}\ll 1$, synchronous rotation is
maintained throughout the subsequent mass transfer episode which starts at
$t=t_{\mathrm{RLOF}}+5\times 10^6$ yr (Fig.~\ref{alpha_self2}c).

Upon He-core ignition, the final mass of the osculating primary is
marginally less massive ($\approx 0.9 M_{\odot}$) than in the classical
models ($1.0 M_{\odot}$), the total system mass being held constant. As
with the case B model, however, the osculating scheme gives a final orbital
period which is a factor of 5.5 smaller ($\approx$ 6~d) than the classical
scheme ($\approx$ 33~d, Fig. \ref{multi2}d).

As for the case B model, $\mathcal{F}_{\mathrm{stream}}^{\mathrm{comp}}$ is
responsible for the shorter post-mass transfer orbital period. The
long-dashed blue curve in Fig.~\ref{multi2} presents the evolution if the
$\mathcal{F}_{\mathrm{stream}}$ terms are neglected. When $q>1$, the
orbital separation slightly increases (Fig.~\ref{multi2}d) because of the
dominant, positive contribution from the $\omega r_{\mathrm{acc}}\cos\psi$
term in Eq.~(\ref{eq:Ty}). The expansion of the orbit reduces the amount
that the primary star over-fills its Roche lobe, giving a correspondingly
smaller mass loss rate peaking at about $2\times 10^{-7}$ \myr
(Fig.~\ref{multi2}c). For a given $t-t_{\mathrm{RLOF}}$, the primary is
therefore more massive than found by the other models, and its
core-hydrogen burning timescale is, in turn, shorter. By consequence, both
shell-hydrogen burning and core-helium burning commence sooner in the
binary's evolution and so the duration of mass transfer is shorter.

Eventually, the secondary will fill its Roche lobe as it evolves off the
main sequence, before He-core burning has ceased in the primary. This
second mass transfer episode starts when the secondary's radius is 17
$R_{\odot}$ for the osculating scheme and 53 $R_{\odot}$ for the classical
model. As for the case B system, we expect common envelope evolution to
follow because of the extreme mass ratio.

%

\section{Discussion} \label{sec:discussion}

\subsection{Consequences of the osculating scheme on the orbital evolution}

Our simulations show that the osculating scheme yields a significantly
shorter post-mass transfer orbital period than the classical
formalism. Alternatively expressed, to obtain an Algol with a given orbital
period, the osculating prescription requires a longer initial
period. Consequently, the progenitor primary may fill its Roche lobe once
it has already developed a deep surface convection zone near the base of
the giant branch. \citet{1988ApJ...334..357T} suggested this was the case
for a number of observed Algols, for example TW Dra and AR Mon. Mass
transfer would therefore proceed on the dynamical - rather than the thermal
- timescale, possibly causing common envelope evolution. They proposed,
however, that such a fate can be avoided if the primary loses sufficient
mass via an enhanced stellar wind \citep[companion-reinforced attrition
  process;][]{1988MNRAS.231..823T} to reduce the mass ratio close to unity,
before RLOF starts.


We currently consider conservative evolution, which may be a reasonable
assumption during the slow mass transfer phase. \citet{2008A&A...487.1129V}
suggested that non-conservative mass transfer in Algols is triggered by a
hotspot located at the secondary's surface, or at the edge of an accretion
disc. They further found that this mechanism typically operates during the
rapid mass transfer phase, and becomes quiescent within the slow
regime. Other suggested mechanisms - albeit poorly studied - include mass
escaping through the outer Lagrangian $\mathcal{L}_{3}$ point
\citep{2007ARep...51..836S} or by bipolar jets
\citep{2007A&A...463..233A}. The removal of orbital angular momentum via
systemic mass losses was also invoked by \citet{1985A&A...142..367D} and
\citet{1994A&A...291..786D} to explain the observed orbital periods of TV
Cas, $\beta$ Lyrae and SV Cen. For TV Cas, for example, the authors
estimated that 80 per cent of the transferred mass is ejected from the
system, removing about 40 per cent of the progenitor's orbital angular
momentum. However, given the effect of $\mathcal{F}_{\mathrm{stream}}$ on
the orbital evolution, we argue that a fraction of the estimated orbital
angular momentum loss can be attributed to the gravitational interaction
between the stars and the mass transfer stream. Therefore, the amount of
orbital angular momentum carried by the expelled mass may be lower than
quoted by \cite{1985A&A...142..367D}. We will consider the impact of
non-conservative evolution in a future investigation.

Our simulations indicate that, contrarily to what is usually assumed, the
primary does not always rotate synchronously throughout mass exchange. As
shown in Figs.~\ref{alpha_self} and \ref{alpha_self2}, the primary is
rotating significantly sub-synchronously during the rapid phase. Only once
the mass transfer rate decelerates and convection develops in the surface
layers, are tides effective enough to re-synchronize the primary.

Unfortunately, published spin rates of the primary during the rapid phase
are, to the best of our knowledge, not available, but this is likely a
result of the short duration of this phase (between about ${10}^{5}$ to
$10^{6}$ years in our models). Existing studies of rapid mass transfer
systems, such as UX Mon \citep{2011A&A...528A.146S}, {\it assume} that the
primary is synchronous, based upon the expectations that tides are always
efficient enough to maintain synchronism, although this has never been
proven observationally. On the other hand, evidence for {\it synchronous}
primaries where the mass ratio has reversed are relatively abundant, such
as the eclipsing $\delta$ Scuti star KIC 10661783
\citep{2013A&A...557A..79L}, TX Uma \citep{2011MNRAS.415.2238G}, KZ Pav
\citep{2010MNRAS.407..497S} and RX Cas \citep{1989A&A...215..272A}.

\subsection{Contact evolution}


The enhanced orbital contraction that self-accretion generates, combined
with expanding radiative secondaries in response to mass accretion
\citep[e.g.][]{1977PASJ...29..249N}, could lead to more contact
systems. The situation may be particularly severe for case A systems when
self-accretion operates even after the mass ratio has reversed. Indeed,
even though the classical scheme predicts orbital expansion when $q<1$,
self-accretion still causes the orbit to contract down to $q \simeq 0.8$ in
our case A simulation. However, contact may be avoided if, for a given
initial period, the mass ratio is initially close to unity. In this
configuration, the mass ratio is reversed sooner, limiting the orbital
shrinkage.

The spin-up of the secondary, as it accretes angular momentum from the
transferred material \citep{1981A&A...102...17P}, may also lead to a
contact binary. \citet{2007ApJ...660.1624S} demonstrated that the Roche
lobe radius of a super-synchronously rotating star will be smaller than the
value determined using the standard \citet{1983ApJ...268..368E}
formula. So, if we also consider the impact of the secondary's rotation on
its Roche radius (as should be done), we indeed find that all our models
enter a contact configuration during the rapid phase. However, magnetic
braking may prevent significant spin-up of the secondary
\citep{2010MNRAS.406.1071D,2013A&A...557A..40D} although, in light of the
expanding radiative secondary, these mechanisms may only delay the onset of
contact.

Contact evolution requires detailed modelling of energy transport between
the stars within their common envelope
\citep[e.g.][]{1977ApJ...215..851W,2009MNRAS.397..857S}, which is beyond
the scope of the present paper. Nonetheless, observations of the contact
systems with early spectral-type stars, such as LY Aur
\citep{2014NewA...26..112Z}, V382 Cyg and TU Mus
\citep{2007MNRAS.380.1599Q} indicate that their evolution is similar to
that of a semi-detached Algol, but much shorter-lived. The authors suggest
that the contact configuration was most likely triggered during a rapid
case A mass transfer phase, and that the observed period increase is caused
by mass transfer from the less massive to the more massive star. They also
expect that the increasing orbital separation will break the contact
configuration, giving a semi-detached Algol, suggesting that not all
contact systems necessarily merge.

Further complications arise from the fact that the rotation rate of
each star is different. The concept of the Roche model, which assumes that
both stars rotate synchronously with the orbit, is therefore incorrect. As
outlined by \citet{1977A&A....54..877V}, asynchronous rotation of both
components gives Roche lobes that do not necessarily coincide at a common
inner-Lagrangian point. This situation may greatly complicate the mass flow
structure between the stars, possibly leading to non-conservative
evolution. We appeal to smooth particle hydrodynamical (SPH) simulations to
investigate this in further detail.

\subsection{Physical considerations}

\subsubsection{Stellar rotation}
\label{sec:rot}

By adopting the solid-body approximation, we assume that torques will spin
up or spin down each star {\it as a whole}. However, it is more likely that
only the outer-most layers will be affected, thereby triggering
differential rotation. Subsequently, angular momentum is re-distributed
within the stellar interior via meridional circulation and shear
instabilities \citep[see][for a review]{2009pfer.book.....M}.

\citet{2013A&A...556A.100S} studied the affect of angular momentum
transport within a differentially rotating 15 $M_{\odot}$ primary, with a
10 $M_{\odot}$ companion as the primary evolved from the ZAMS until the
onset of RLOF. They found that meridional circulation always counteracts
the impact of tides; spinning up the surface layers when tides spin them
down and vice versa, increasing the time for the star to rotate
synchronously with the orbit. Consequently, some of their models commence
RLOF before the primary is synchronised. In addition, differential rotation
significantly enhances nitrogen abundances at the stellar surface.


We can speculate that significant differential rotation will occur while
our donors become sub-synchronous, with meridional circulation initially
opposing the spin-down triggered by rapid mass loss, and then the
subsequent tidal forces which act to re-synchronise the donor. In turn,
this may affect both the duration of the self-accretion phase and the value
of $\alpha_{\mathrm{self}}$.

\subsubsection{Tides}
\label{sec:tides}

\citet{1987ApJ...322..856T,1988ApJ...324L..71T} proposed an alternative
  theory that invokes large-scale hydrodynamical flows as a means to
  dissipate kinetic energy, which are more efficient than the dynamical
  tide model described by \citet{1977A&A....57..383Z}. Indeed, the
  synchronization and circularisation timescales between the two approaches
  vary by up to three orders of magnitude \citep{2010MNRAS.401..257K}.

  We can speculate on the impact of Tassoul's formalism on our calculations
  as follows. If $\tau_{\mathrm{sync}}$ is 1000 times shorter, then for our
  case B system $\tau_{\mathrm{sync}}<\tau_{R}$, i.e. tides keep the donor
  in synchronous rotation by the time RLOF starts. Additionally, for both
  the case A and case B models, an inspection of Figs. \ref{alpha_self} and
  \ref{alpha_self2} shows that we would have
  $\tau_{\mathrm{sync}}<\tau_{\dot{M}}$. Hence, even during the fast phase,
  mass loss would not be sufficiently rapid to spin down the donor star to
  sub-synchronous rates and, in turn, self-accretion would not be
  triggered. In this case, the orbital evolution will correspond to the
  short-dashed green curves given in Figs. \ref{multi} and \ref{multi2},
  where we neglected the $\mathcal{T}_{\mathrm{self}}$ term. Therefore,
  using Tassoul's formalism will not significantly affect our results,
  namely that our osculating scheme still yields much shorter orbital
  periods.

  However, the hydrodynamical model has since been criticised on
  theoretical grounds by \citet{1992A&A...259..581R} and
  \citet{1997ApJ...474..760R} \citep[but see][for a
    counter-argument]{1997ApJ...481..363T}, while observations attempting
  to constrain the mechanism underpinning tidal interactions are
  inconclusive. On the one hand, \citet{1995A&A...299..724C} and
  \citet{1997A&A...318..187C} found that both the Zahn and Tassoul
  formalisms can adequately account for the observed eccentricities and
  rotational velocities of early-type eclipsing binaries. On the other
  hand, using an updated sample of such binaries,
  \citet{2007MNRAS.382..356K,2010MNRAS.401..257K} found that Tassoul's
  theory is in contradiction with observations, which are better reproduced
  by Zahn's formalism. However, \citet{2005ApJ...620..970M} derived, for
  stellar populations with a range of ages, the tidal circularization
  period (i.e the orbital period at which the binary orbit circularises at
  the age of the population). Their results indicated that, at a given age,
  the observed circularisation period is larger than the predicted value
  from the dynamical tide model, suggesting that it is too
  inefficient. Clearly, more observational and theoretical work is required
  in this area.

\subsubsection{Self-accretion}
\label{sec:self_accretion}

The phenomenon of self-accretion has also been found in the studies of
\citet{1964AcA....14..231K} and \citet{2010ApJ...724..546S}, who use the
same ballistic approach, which assumes that no collisions between particles
occur. Clearly, this is not realistic, as there are indeed multiple
collisions along the trajectory.

Nonetheless, our results are in qualitative agreement with the SPH
calculations of \citet{1993A&A...280..525B}, who find that below some
critical rotational velocity, the entire stream is deflected towards the
primary. For larger spin rates, material falls onto both
components. However, their study focused on the impact of asynchronous
rotation on disc formation around the secondary, and they did not quantify
self-accretion. We hope our work will motivate future SPH studies in this
area.

\section{Summary and conclusions} \label{sec:conclusions}

We use our stellar binary evolution code BINSTAR to calculate the evolution
of Algol systems using the theory of osculating orbital elements. By
calculating the ballistic trajectories of ejected particles from the mass
losing star (the donor), we determine the change of linear momentum of each
star, and the gravitational perturbation applied to the stars by the mass
transfer stream. As a consequence of the latter, the osculating formalism
predicts significantly shorter post-mass transfer orbital periods,
typically by a factor of 4, than the widely applied classical scheme.

Also contrary to widely held belief, the donor star does not remain in
synchronous rotation with the orbital motion throughout mass exchange. The
initially rapid mass ejection spins down the donor on a shorter timescale
than the tidal synchronization timescale, enforcing sub-synchronous
rotation and causing about 15 to 20 per cent of the ejected material
to fall back onto the donor during these episodes of
self-accretion. Self-accretion, combined with the sink of orbital angular
momentum that mass transfer provides, may lead to the formation of more
contact binary systems.

While we have mainly focused on conservative Algol evolution, the
osculating prescription clearly applies to all varieties of interacting
binaries. In the future, we will apply our osculating formalism to
investigate the evolution of eccentric systems.
 
\begin{acknowledgements}

PJD acknowledges financial support from the FNRS Research Fellowship -
Charg{\'{e}} de Recherche. LS is a FNRS Research Associate. We thank the
anonymous referee whose constructive comments helped to improve the quality
of the manuscript.
     
\end{acknowledgements}




\appendix

\section{The perturbing forces $\mathcal{S}_{\mathrm{self}}$ and
  $\mathcal{T}_{\mathrm{self}}$ } 
\label{self_accretion}

\begin{figure}
  \begin{center}
    \includegraphics[scale=0.4]{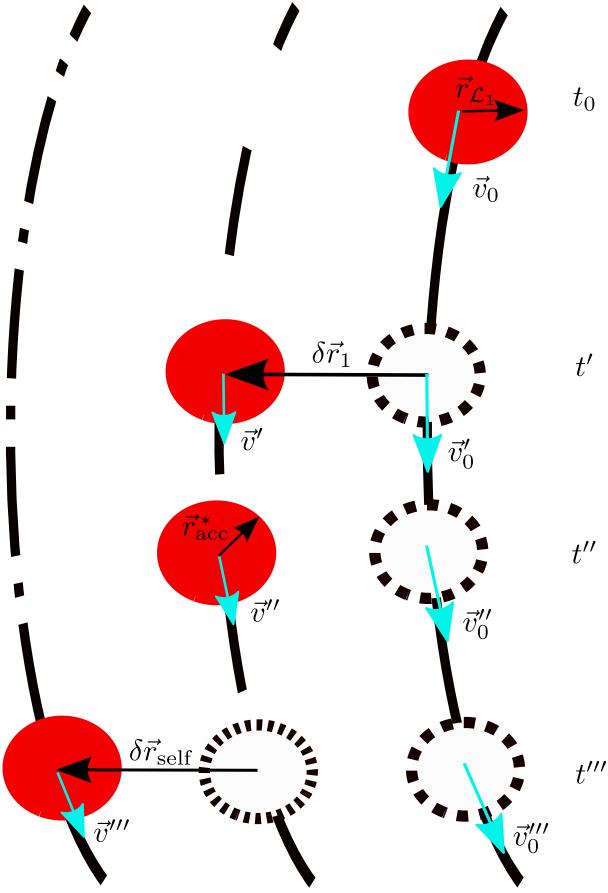}
    \caption{Illustration of the self-accretion process. At time $t=t_{0}$,
      the donor star moves along its orbital path (solid black curve), with
      an orbital velocity $\vec{v}_{0}$ (cyan arrow). At $t^{\prime}$, a
      particle of mass $\delta M_{1,\mathrm{ej}}^{\mathrm{self}}$ is
      ejected from the inner-Lagrangian point located at
      $\vec{r}_{\mathcal{L}_{1}}$ with respect to the donor's mass
      centre. The ejection shifts the centre of mass by $\delta
      \vec{r}_{1}$, and the donor follows a new orbit (long dashed curve)
      with velocity $\vec{v}^{\prime}$. At $t^{\prime\prime}$ just before
      the particle is re-accreted (at $\vec{r}_{\mathrm{acc}}^{*}$), the
      orbital velocity is $\vec{v}^{\prime\prime}$. Subsequently,
      self-accretion shifts the donor's mass centre by $\delta
      \vec{r}_{\mathrm{self}}$, and it follows the orbit indicated by the
      dot-dashed curve, with a velocity $\vec{v}^{\prime\prime\prime}$. The
      dashed circles represent the locations of the donor if no mass
      ejection had taken place, while the dotted circle indicates the
      donor's location had no self-accretion occurred.}
    \label{self_accretion_fig}
  \end{center}
\end{figure}

Consider a primary star of mass $M_{1}$. At time $t=t_{0}$, it has an
orbital velocity $\vec{v}_{1}=\vec{v}_{0}$, and an orbital angular velocity
$\vec{\omega}$. At a later time $t=t^{\prime}$, a particle is ejected from
the primary at the inner-Lagrangian point, located at a distance
$r_{\mathcal{L}_{1}}$ from the primary's mass centre (see
Fig. \ref{self_accretion_fig}), and so the primary's mass becomes
$M_{1}+\delta{M}_{\mathrm{1,ej}}^{\mathrm{self}}$, where
$\delta{M}_{\mathrm{1,ej}}^{\mathrm{self}}<0$. The particle's absolute
velocity is $\vec{W}^{\prime}_{\mathrm{ej}}$. As a result, the centre of
mass is shifted by
\begin{equation}
\delta{\vec{r}_{1}}=\frac{\delta{M}_{1,\mathrm{ej}}^{\mathrm{self}}}{M_{1}}\vec{r}_{\mathcal{L}_{1}}
\label{dr}
\end{equation} 
with respect to its unperturbed location, and its new orbital velocity is
$\vec{v}_{1}^\prime$. During a time interval $\delta{t}$, the change in the
primary's orbital velocity is \citep[see][for further
  details]{1969Ap&SS...3...31H,2007ApJ...667.1170S}
\begin{eqnarray}
  \frac{\vec{v}_{1}^{\prime}-\vec{v}_{1}}{\delta t} & = & 
  \frac{\vec{Q}^{\prime}_{1}-\vec{Q}_{1}}{\delta t}\frac{1}{M_{1}}+\frac{\delta
    M_{1,\mathrm{ej}}^{\mathrm{self}}}{\delta
    t}\frac{1}{M_{1}}(\vec{V}_{\mathrm{ej}}+\vec{\omega}^{\prime}\vec{\times}\vec{r}_{\mathcal{L}_{1}})\nonumber \\
  & & +\frac{1}{M_{1}}\frac{\delta M_{1,\mathrm{ej}}^{\mathrm{self}}}{(\delta t)^{2}}\vec{r}_{\mathcal{L}_{1}},
\label{dV_1}
\end{eqnarray}
where $\vec{Q}_{1}$ is the primary's momentum, primed quantities indicate
values at time $t^{\prime}$, and
$\vec{V}_{\mathrm{ej}}=\vec{W}_{\mathrm{ej}}-\vec{v}_{1}$ is the relative velocity of
the ejected material with respect to the primary's mass centre.

At self-accretion, the primary accretes the particle of mass
$\delta{M}_{1,\mathrm{acc}}^{\mathrm{self}}=-\delta{M}_{1,\mathrm{ej}}^{\mathrm{self}}>0$. Just
before self-accretion occurs at time $t=t^{\prime\prime}$, the momentum of
the primary and ejected particle, $\vec{Q}^{\prime\prime}_{1}$, is
\begin{equation}
\vec{Q}_{1}^{\prime\prime}=(M_{1}-\delta M_{1,\mathrm{acc}}^{\mathrm{self}})\vec{v}_{1}^{\prime\prime}+\delta M_{1,\mathrm{acc}}^{\mathrm{self}}\vec{W}_{\mathrm{acc}}
\label{Q_pprime}
\end{equation}
where $\vec{W}_{\mathrm{acc}}$ is the absolute velocity of the
self-accreted particle. The orbital velocity, $\vec{v}_{1}^{\prime\prime}$,
is the sum of the non-perturbed orbital velocity at time
$t^{\prime\prime}$, $\vec{v}_{0}^{\prime\prime}$, and the perturbation
to the velocity because of the original ejection episode, so
\begin{equation}
  \vec{v}_{1}^{\prime\prime}=\vec{v}_{0}^{\prime\prime}+(\vec{\omega}^\prime+\delta{\vec{\omega}})\vec{\times}
      {\delta{\vec{r}}_{1}}=\vec{v}_{0}^{\prime\prime}+\frac{\delta{M}_{1,\mathrm{ej}}^{\mathrm{self}}}{M_{1}}(\vec{\omega}^\prime
      \vec{\times} \vec{r}_{\mathcal{L}_{1}}),
\label{v_pprime}
\end{equation}
where we have used Eq.~(\ref{dr}) and ignored terms larger than
first-order. Inserting Eq.~(\ref{v_pprime}) into Eq.~(\ref{Q_pprime}) gives
\begin{equation}
\vec{Q}_{1}^{\prime\prime}= M_{1}\vec{v}_{0}^{\prime\prime}+\delta
M_{1,\mathrm{ej}}^{\mathrm{self}}(\vec{\omega}^{\prime}\vec{\times}\vec{r}_{\mathcal{L}_{1}})-\delta
M_{1,\mathrm{acc}}^{\mathrm{self}}\vec{v}_{0}^{\prime\prime}+\delta M_{1,\mathrm{acc}}^{\mathrm{self}}\vec{W}_{\mathrm{acc}}.
\label{Q_pprime_2}
\end{equation}
The particle is self-accreted at time $t^{\prime\prime\prime}$, at a
location $\vec{r}_{\mathrm{acc}}^{*}$ with respect to the primary's mass
centre. This shifts the mass centre by an amount
\begin{equation}
\delta{\vec{r}_{\mathrm{self}}}=\frac{\delta{M}_{1,\mathrm{acc}}^{\mathrm{self}}}{M_{1}}\vec{r}^{*}_{\mathrm{acc}}
\label{dr_self}
\end{equation}
and the primary's momentum is now
\begin{equation}
\vec{Q}^{\prime\prime\prime}=M_{1}\vec{v}_{1}^{\prime\prime\prime}.
\label{Q_ppprime}
\end{equation}
The new orbital velocity, $\vec{v}_{1}^{\prime\prime\prime}$, is the sum of
the unperturbed orbital velocity, $v^{\prime\prime\prime}_{0}$, and the
perturbations to the velocity arising from the ejection and
self-accretion processes, i.e.
\begin{eqnarray}
  \vec{v}_{1}^{\prime\prime\prime}& = & \vec{v}_{0}^{\prime\prime\prime}+\frac{\delta{M}_{1,\mathrm{ej}}^{\mathrm{self}}}{M_{1}}(\vec{\omega}^\prime
  \vec{\times}
  \vec{r}_{\mathcal{L}_{1}})+(\vec{\omega}^{\prime\prime}+\delta{\vec{\omega}})\vec{\times}
  \delta{\vec{r}_{\mathrm{self}}}\nonumber \\ 
& = & \vec{v}_{0}^{\prime\prime\prime}+\frac{\delta{M}_{1,\mathrm{ej}}^{\mathrm{self}}}{M_{1}}(\vec{\omega}^\prime
  \vec{\times}
  \vec{r}_{\mathcal{L}_{1}}) +
  \frac{\delta{M}_{1,\mathrm{acc}}^{\mathrm{self}}}{M_{1}}(\vec{\omega}^{\prime\prime}
  \vec{\times} \vec{r}_{\mathrm{acc}}^{*})
\label{v_ppprime}
\end{eqnarray}
where we have used Eq.~(\ref{dr_self}), and once again ignored terms higher
than first-order. Inserting Eq.~(\ref{v_ppprime}) into
Eq.~(\ref{Q_ppprime}) yields
\begin{equation}
\vec{Q}_{1}^{\prime\prime\prime}=M_{1}\vec{v}_{0}^{\prime\prime\prime}+\delta{M_{1,\mathrm{ej}}^{\mathrm{self}}}(\vec{\omega}^{\prime}\vec{\times}\vec{r}_{\mathcal{L}_{1}})+\delta M_{1,\mathrm{acc}}^{\mathrm{self}}(\vec{\omega}^{\prime\prime}\vec{\times}\vec{r}_{\mathrm{acc}}^{*}).
\label{Q_ppprime_2}
\end{equation}
Taking the difference between Eqs.~(\ref{Q_ppprime_2}) and
(\ref{Q_pprime_2}), dividing the result by $\delta{t}$ and using the fact
that
$\delta{M}_{1,\mathrm{ej}}^{\mathrm{self}}=-\delta{M}_{1,\mathrm{acc}}^{\mathrm{self}}$
yields
\begin{eqnarray}
\frac{\vec{v}_{0}^{\prime\prime\prime}-\vec{v}_{0}^{\prime\prime}}{\delta
  t} & = &
\frac{1}{M_{1}}\frac{\vec{Q}_{1}^{\prime\prime\prime}-\vec{Q}_{1}^{\prime\prime}}{\delta
  t}-\frac{\delta M_{1,\mathrm{acc}}^{\mathrm{self}}}{\delta
  t}\frac{1}{M_{1}}(\vec{\omega}^{\prime\prime}\vec{\times}\vec{r}_{\mathrm{acc}}^{*})\nonumber
\\ & - &\frac{\delta M_{1,\mathrm{acc}}^{\mathrm{self}}}{\delta
  t}\frac{1}{M_{1}}\vec{v}_{0}^{\prime\prime}+\frac{\delta
  M_{1,\mathrm{acc}}^{\mathrm{self}}}{\delta t}\frac{1}{M_{1}}\vec{W}_{\mathrm{acc}}.
\label{dV_2}
\end{eqnarray}
Following \citet{2007ApJ...667.1170S}, the absolute acceleration of the
primary's mass centre is the sum of Eq.~(\ref{dV_2}), the relative
acceleration of the primary's mass centre,
$\frac{\dot{M}_{1,\mathrm{acc}}^{\mathrm{self}}}{(\delta
  t)^{2}}\frac{1}{M_{1}}$, and the Coriolis acceleration,
$\frac{1}{M_{1}}\frac{\delta{M}_{1,\mathrm{acc}}^{\mathrm{self}}}{\delta
  t}2(\vec{\omega}^{\prime\prime}\vec{\times} \vec{r}_{\mathrm{acc}}^{*})$ to give
\begin{eqnarray}
\frac{\vec{v}_{1}^{\prime\prime\prime}-\vec{v}_{1}^{\prime\prime}}{\delta
  t}& = & \frac{1}{M_{1}}\frac{\vec{Q}_{1}^{\prime\prime\prime}-\vec{Q}_{1}^{\prime\prime}}{\delta
  t}+\frac{\delta M_{1,\mathrm{acc}}^{\mathrm{self}}}{\delta
  t}\frac{1}{M_{1}}[(\vec{\omega}^{\prime\prime}\vec{\times}
  \vec{r}_{\mathrm{acc}}^{*})+\vec{V}_{\mathrm{acc}}]\nonumber \\ & & +\frac{1}{M_{1}}\frac{\delta
  M_{1,\mathrm{acc}}^{\mathrm{self}}}{(\delta t)^{2}},
\label{dV_self}
\end{eqnarray}
where
$\vec{V}_{\mathrm{acc}}=\vec{W}_{\mathrm{acc}}-\vec{v}_{0}^{\prime\prime}$
is the relative velocity of the self-accreted particle with respect to the
primary's mass centre.

Summing Eqs.~(\ref{dV_self}) and (\ref{dV_1}) gives the acceleration of the
primary's mass centre from both the ejection and self-accretion
process. In the limit $\delta t\rightarrow 0$, and remembering that
$\delta{M}_{1,\mathrm{acc}}^{\mathrm{self}}=-\delta{M}_{1,\mathrm{ej}}^{\mathrm{self}}$,
then the acceleration of the primary's mass centre is
\begin{eqnarray}
\frac{\mathrm{d}^{2}\vec{R}_{1}}{\mathrm{d}t^{2}}& = &\frac{\vec{F}_{1}}{M_{1}}+\frac{\dot{M}_{1,\mathrm{ej}}^{\mathrm{self}}}{M_{1}}[(\vec{V}_{\mathrm{ej}}-\vec{V}_{\mathrm{acc}})+(\vec{\omega}^{\prime}\vec{\times}
  \vec{r}_{\mathcal{L}_{1}}-\vec{\omega}^{\prime\prime}\vec{\times}
  \vec{r}_{\mathrm{acc}}^{*})]\nonumber \\ & & +\frac{\ddot{M}_{1,\mathrm{ej}}^{\mathrm{self}}}{M_{1}}(\vec{r}_{\mathcal{L}_{1}}-\vec{r}_{\mathrm{acc}}^{*}),
\label{dR1_dt}
\end{eqnarray}
where $\vec{R}_{1}$ is the position vector of the primary with respect to
an inertial reference frame, and $\vec{F}_{1}$ is the sum of all external
forces acting on the primary, which writes as
\begin{equation}
\vec{F}_{1}=-\frac{GM_{1}M_{2}}{r^{2}}\frac{\vec{R}_{1}}{R_{1}}+\vec{f}_{1},
\label{F1}
\end{equation}
where $\vec{r}=\vec{R}_{2}-\vec{R}_{1}$, $\vec{R}_{2}$ is the position
vector of the secondary, and $\vec{f}_{1}$ is the force acting on the
primary via the matter stream. Since the secondary is not accreting, its
equation of motion is
\begin{equation}
\frac{\mathrm{d}^{2}\vec{R}_{2}}{\mathrm{d}t^{2}}=-\frac{GM_{1}}{r^{2}}\frac{\vec{R}_{2}}{R_{2}}+\frac{\vec{f}_{2}}{M_{2}},
\label{dR2_dt}
\end{equation}
where $\vec{f}_{2}$ is the force acting on the secondary by the
accretion stream. Subtracting Eq.~(\ref{dR1_dt}) from Eq.~(\ref{dR2_dt})
gives the equation of motion of the secondary with respect to the primary,
which is
\begin{eqnarray}
\frac{\mathrm{d}^{2}\vec{r}}{\mathrm{d}t^{2}} & = & -\frac{G(M_1+M_{2})}{r^{3}}\vec{r}+\frac{\vec{f}_{2}}{M_{2}}-\frac{\vec{f}_{1}}{M_{1}}-\frac{\dot{M}_{1,\mathrm{ej}}^{\mathrm{self}}}{M_{1}}[(\vec{V}_{\mathrm{ej}}-\vec{V}_{\mathrm{acc}})\nonumber
  \\ & & +(\vec{\omega}^{\prime}\vec{\times}
  \vec{r}_{\mathcal{L}_{1}}-\vec{\omega}^{\prime\prime}\vec{\times}
  \vec{r}_{\mathrm{acc}}^{*})]-\frac{\ddot{M}_{1,\mathrm{ej}}^{\mathrm{self}}}{M_{1}}(\vec{r}_{\mathcal{L}_{1}}-\vec{r}_{\mathrm{acc}}^{*}),
\label{dr2_dt2}
\end{eqnarray}
which takes the form
\begin{equation}
\frac{\mathrm{d}^{2}\vec{r}}{\mathrm{d}t^{2}}=-\frac{G(M_1+M_{2})}{r^{3}}\vec{r}+\mathcal{S}_{\mathrm{self}}\
\hat{e}_{r}+\mathcal{T}_{\mathrm{self}}\
\hat{e}_{t}.
\label{d2r_dt2_2}
\end{equation}
Here, $\hat{e}_{r}$ is a unit vector pointing along $\vec{r}$, and
$\hat{e}_{t}$ is a unit vector perpendicular to
$\hat{e}_{r}$ in the direction of the orbital motion. Taking the
dot product of Eq.~(\ref{d2r_dt2_2}) with $\hat{e}_{r}$ and
$\hat{e}_{t}$ respectively, yields
\begin{eqnarray}
\mathcal{S}_{\mathrm{self}} & =
&\frac{f_{2,r}}{M_{2}}-\frac{f_{1,r}}{M_{1}}-\frac{\ddot{M}_{1,\mathrm{ej}}^{\mathrm{self}}}{M_{1}}(r_{\mathcal{L}_{1}}-r_{\mathrm{acc}}^{*}\cos\psi^{*})\nonumber\\ & &
-\frac{\dot{M}_{1,\mathrm{ej}}^{\mathrm{self}}}{M_{1}}(V_{\mathrm{ej},r}-V_{\mathrm{acc},r}+\omega^{\prime\prime}r_{\mathrm{acc}}^{*}\sin\psi^{*}),
\end{eqnarray}
and
\begin{eqnarray}
\mathcal{T}_{\mathrm{self}}& =
&\frac{f_{2,t}}{M_{2}}-\frac{f_{1,t}}{M_{1}}-\frac{\ddot{M}_{1,\mathrm{ej}}^{\mathrm{self}}}{M_{1}}r_{\mathrm{acc}}^{*}\sin\psi^{*}\nonumber\\ & & -\frac{\dot{M}_{1,\mathrm{ej}}^{\mathrm{self}}}{M_{1}}(V_{\mathrm{ej},t}-V_{\mathrm{acc},t}+\omega^{\prime}r_{\mathcal{L}_{1}}-\omega^{\prime\prime}r_{\mathrm{acc}}^{*}\cos\psi^{*}),
\end{eqnarray}
which are the same as Eq.~(\ref{eq:Sx_self}) and (\ref{eq:Ty_self}), noting
that $\omega^{\prime}=\omega^{\prime\prime}=\omega$ for circular
orbits. The quantity $\psi^{*}$ is the angle between $\hat{e}_{r}$ and the
impact site, and the subscripts `$r$' and `$t$' indicate components along
$\hat{e}_{r}$ and $\hat{e}_{t}$ respectively.

\section{Torque arising from mass transfer, $\dot{J}_{\mathrm{MT}}$}
\label{app:Jdot_RLOF}

Consider a primary star of mass $M_{1}$, and a secondary of mass $M_{2}$,
separated by a distance $r$. They respectively orbit the common centre of
mass with an orbital velocity $\vec{v}_{1}$ and $\vec{v}_{2}$. The velocity
of the secondary with respect to the primary is
$\vec{v}=\vec{v}_{2}-\vec{v}_{1}$, and the orbital angular momentum is
given by
\begin{equation}
J_{\mathrm{orb}}=m[GMa(1-e^{2})]^{1/2}=mrv_{t},
\label{J_orb}
\end{equation}
where $e$ is the eccentricity, $m=M_{1}M_{2}/M$ is the reduced mass,
$M=M_{1}+M_{2}$, and $v_{t}$ is the orbital velocity along
$\hat{e}_{t}$, given by
\begin{equation}
v_{t}=\left[\frac{GM}{a(1-e^{2})}\right]^{1/2}(1+e\cos\nu)
\label{vt}
\end{equation}
and $\nu$ is the true anomaly.
Taking the time derivative of the last equality in Eq.~(\ref{J_orb}) and noting
that $\dot{r}=0$ for an osculating orbit \citep[see,
  e.g.][]{2008A&A...480..797B}, yields
\begin{equation}
\frac{\dot{J}_{\mathrm{orb}}}{J_{\mathrm{orb}}}=\frac{\dot{M}_{1}}{M_{1}}+\frac{\dot{M}_{2}}{M_{2}}-\frac{\dot{M}}{M}+\frac{\dot{v}_{t}}{v_{t}}.
\label{Jdot_J}
\end{equation}
Similarly to Eq.~(\ref{d2r_dt2_2}), the equation of motion of a binary
acted on by perturbing forces $\mathcal{S}$ and $\mathcal{T}$ reads
\begin{equation}
\frac{\mathrm{d}^{2}\vec{r}}{\mathrm{d}t^{2}}=\frac{\mathrm{d}\vec{v}}{\mathrm{d}t}=-\frac{G(M_1+M_{2})}{|r|^{3}}\vec{r}+\mathcal{S}\hat{e}_{r}+\mathcal{T}\hat{e}_{t}.
\label{dv_dt}
\end{equation}
Taking the dot product of Eq.~(\ref{dv_dt}) with $\hat{e}_{t}$
gives
\begin{equation}
\frac{\mathrm{d}v_{t}}{\mathrm{d}t}=\mathcal{T}.
\label{dvt_dt}
\end{equation}
Inserting Eqs.~(\ref{dvt_dt}) and (\ref{vt}) into Eq.~(\ref{Jdot_J}), and
using the first equality in Eq.~(\ref{J_orb}), gives the torque applied
onto the orbit from mass transfer,
\begin{equation}
  \dot{J}_{\mathrm{orb,MT}}=J_{\mathrm{orb}}\left(\frac{\dot{M}_{1}}{M_{1}}+\frac{\dot{M}_{2}}{M_{2}}-\frac{\dot{M}}{M}\right)+m\frac{a(1-e^{2})}{1+e\cos\nu}\mathcal{T}.
\label{J_orb_2}
\end{equation}
The net change of the primary's mass is the sum of mass transferred to the
companion via RLOF, $\dot{M}_{1,\mathrm{ej}}^{\mathrm{comp}}<0$, and mass
ejected by the wind, $\dot{M}_{1,\mathrm{loss}}<0$, i.e.
\begin{equation}
\dot{M}_{1}=\dot{M}_{1,\mathrm{ej}}^{\mathrm{comp}}+\dot{M}_{\mathrm{1,loss}}.
\label{M1dot}
\end{equation}
Similarly, for the secondary
\begin{equation}
\dot{M}_{2}=-\beta\dot{M}_{1,\mathrm{ej}}^{\mathrm{comp}}+\dot{M}_{\mathrm{2,loss}},
\label{M2dot}
\end{equation}
where the first term on the right hand side gives the accretion rate and
$\dot{M}_{2,\mathrm{loss}}$ includes the mass ejected from the system
during non-conservative mass transfer. Substituting Eqs.~(\ref{M1dot}) and
(\ref{M2dot}) into Eq.~(\ref{J_orb_2}) gives
\begin{eqnarray}
  \dot{J}_{\mathrm{orb,MT}} & = &
  \underbrace{J_{\mathrm{orb}}\left[\frac{\dot{M}_{1,\mathrm{ej}}^{\mathrm{comp}}}{M}\left(\frac{1}{q}-\beta{q}\right)\right]+m \frac{a(1-e^{2})}{1+e\cos\nu}\mathcal{T}}_{\dot{J}_{\mathrm{orb,RLOF}}}\nonumber \\ & &
  +\underbrace{J_{\mathrm{orb}}\left(\frac{\dot{M}_{1,\mathrm{loss}}}{M}\frac{1}{q}+\frac{\dot{M}_{2,\mathrm{loss}}}{M}q\right)}_{\dot{J}_{\mathrm{orb,lost}}}
\label{J_orb_3}
\end{eqnarray}
where $q=M_{1}/M_{2}$, $\dot{J}_{\mathrm{orb,RLOF}}$ is the torque acting
on the orbit as a consequence of Roche lobe overflow, while
$\dot{J}_{\mathrm{orb,lost}}<0$ is the torque applied by the material
leaving the system. The corresponding torque applied onto the transferred
mass is just
\begin{equation}
  \dot{J}_{\mathrm{MT}}=-\dot{J}_{\mathrm{orb,MT}}.
\label{Jdot_RLOF_app}
\end{equation}
Using Eqs.~(\ref{J_orb_3}) and (\ref{Jdot_RLOF_app}) with $e=0$ gives
Eq.~(\ref{Jdot_stream}).

Next, we demonstrate the consistency of Eq.~(\ref{J_orb_3}) by showing that
in the classical formalism for conservative mass transfer
$\dot{J}_{\mathrm{orb,RLOF}}=0$. If all material is transferred to the
secondary ($\alpha_{\mathrm{self}}=0$, $\beta=1$), if the stars are treated
as point masses ($r_{\mathcal{L}_{1}}=0$, $r_{\mathrm{acc}}=0$), and if we
neglect the gravitational attraction by the accretion stream
($\vec{f}_{1}=0$, $\vec{f}_{2}=0$), Eq.~(\ref{eq:Ty}) reduces to
\begin{equation}
\mathcal{T}_{\mathrm{comp}}=-\frac{\dot{M}_{1,\mathrm{ej}}^{\mathrm{comp}}}{M_{1}}(qV_{2,t}+V_{1,t}),
\label{T_comp_canon}
\end{equation}
where $\dot{M}_{2,\mathrm{acc}}=-\dot{M}_{1,\mathrm{ej}}^{\mathrm{comp}}$ for conservative
mass transfer. \citet{2007ApJ...667.1170S} and \citet{2008ARep...52..680L}
demonstrated that, if the orbital angular momentum is conserved, in a
circular orbit $V_{1,t}$ and $V_{2,t}$ are related by
\begin{equation}
qV_{2,t}+V_{1,t}=\left(\frac{GM}{a}\right)^{1/2}(1-q).
\label{v1_v2}
\end{equation}
Substituting Eqs.~(\ref{v1_v2}) into Eq.~(\ref{T_comp_canon}), and that
result into Eq.~(\ref{J_orb_2}) gives for a circular orbit
\begin{eqnarray}
  \dot{J}_{\mathrm{orb,MT}} & = & \dot{J}_{\mathrm{orb,RLOF}} = 
  J_{\mathrm{orb}}\frac{\dot{M}_{1,\mathrm{ej}}^{\mathrm{comp}}}{M_{1}}\left(\frac{q}{1+q}\right)\left(\frac{1-q^{2}}{q}\right)\nonumber \\ & &
  -\frac{\dot{M}_{1,\mathrm{ej}}^{\mathrm{comp}}}{M_{1}}m\left(GMa\right)^{1/2}(1-q)
\label{Jdot_orb_canon}
\end{eqnarray}
where we have used $M_{1}/M=q/(1+q)$. Using Eq.~(\ref{J_orb}),
Eq.~(\ref{Jdot_orb_canon}) reduces to zero, as required.

\bibliographystyle{aa}
\bibliography{davis_references}

\begin{thebibliography}{80}
\expandafter\ifx\csname natexlab\endcsname\relax\def\natexlab#1{#1}\fi

\bibitem[{{Ak} {et~al.}(2007){Ak}, {Chadima}, {Harmanec}, {Demircan}, {Yang},
  {Koubsk{\'y}}, {{\v S}koda}, {{\v S}lechta}, {Wolf}, {Bo{\v z}i{\'c}}, {Ru{\v
  z}djak}, \& {Sudar}}]{2007A&A...463..233A}
{Ak}, H., {Chadima}, P., {Harmanec}, P., {et~al.} 2007, \aap, 463, 233

\bibitem[{{Andersen} {et~al.}(1990){Andersen}, {Nordstrom}, \&
  {Clausen}}]{1990A&A...228..365A}
{Andersen}, J., {Nordstrom}, B., \& {Clausen}, J.~V. 1990, \aap, 228, 365

\bibitem[{{Andersen} {et~al.}(1989){Andersen}, {Pavlovski}, \&
  {Piirola}}]{1989A&A...215..272A}
{Andersen}, J., {Pavlovski}, K., \& {Piirola}, V. 1989, \aap, 215, 272

\bibitem[{{Belvedere} {et~al.}(1993){Belvedere}, {Lanzafame}, \&
  {Molteni}}]{1993A&A...280..525B}
{Belvedere}, G., {Lanzafame}, G., \& {Molteni}, D. 1993, \aap, 280, 525

\bibitem[{{Bona{\v c}i{\'c} Marinovi{\'c}} {et~al.}(2008){Bona{\v c}i{\'c}
  Marinovi{\'c}}, {Glebbeek}, \& {Pols}}]{2008A&A...480..797B}
{Bona{\v c}i{\'c} Marinovi{\'c}}, A.~A., {Glebbeek}, E., \& {Pols}, O.~R. 2008,
  \aap, 480, 797

\bibitem[{{Chen} {et~al.}(2013){Chen}, {Han}, {Deca}, \&
  {Podsiadlowski}}]{2013MNRAS.tmp.1615C}
{Chen}, X., {Han}, Z., {Deca}, J., \& {Podsiadlowski}, P. 2013, \mnras

\bibitem[{{Claret} \& {Cunha}(1997)}]{1997A&A...318..187C}
{Claret}, A. \& {Cunha}, N.~C.~S. 1997, \aap, 318, 187

\bibitem[{{Claret} {et~al.}(1995){Claret}, {Gimenez}, \&
  {Cunha}}]{1995A&A...299..724C}
{Claret}, A., {Gimenez}, A., \& {Cunha}, N.~C.~S. 1995, \aap, 299, 724

\bibitem[{{Davis} {et~al.}(2013){Davis}, {Siess}, \&
  {Deschamps}}]{2013A&A...556A...4D}
{Davis}, P.~J., {Siess}, L., \& {Deschamps}, R. 2013, \aap, 556, A4

\bibitem[{{De Greve} \& {Linnell}(1994)}]{1994A&A...291..786D}
{De Greve}, J.~P. \& {Linnell}, A.~P. 1994, \aap, 291, 786

\bibitem[{{De Greve} {et~al.}(1985){De Greve}, {Packet}, \& {de
  Landtsheer}}]{1985A&A...142..367D}
{De Greve}, J.~P., {Packet}, W., \& {de Landtsheer}, A.~C. 1985, \aap, 142, 367

\bibitem[{{Dervi{\c s}o{\v g}lu} {et~al.}(2010){Dervi{\c s}o{\v g}lu}, {Tout},
  \& {Ibano{\v g}lu}}]{2010MNRAS.406.1071D}
{Dervi{\c s}o{\v g}lu}, A., {Tout}, C.~A., \& {Ibano{\v g}lu}, C. 2010, \mnras,
  406, 1071

\bibitem[{{Deschamps} {et~al.}(2013){Deschamps}, {Siess}, {Davis}, \&
  {Jorissen}}]{2013A&A...557A..40D}
{Deschamps}, R., {Siess}, L., {Davis}, P.~J., \& {Jorissen}, A. 2013, \aap,
  557, A40

\bibitem[{{Edwards} \& {Pringle}(1987)}]{1987MNRAS.229..383E}
{Edwards}, D.~A. \& {Pringle}, J.~E. 1987, \mnras, 229, 383

\bibitem[{{Eggleton}(1983)}]{1983ApJ...268..368E}
{Eggleton}, P.~P. 1983, \apj, 268, 368

\bibitem[{{Flannery}(1975)}]{1975MNRAS.170..325F}
{Flannery}, B.~P. 1975, \mnras, 170, 325

\bibitem[{{Geller} \& {Mathieu}(2011)}]{2011Natur.478..356G}
{Geller}, A.~M. \& {Mathieu}, R.~D. 2011, \nat, 478, 356

\bibitem[{{Glazunova} {et~al.}(2011){Glazunova}, {Mkrtichian}, \&
  {Rostopchin}}]{2011MNRAS.415.2238G}
{Glazunova}, L.~V., {Mkrtichian}, D.~E., \& {Rostopchin}, S.~I. 2011, \mnras,
  415, 2238

\bibitem[{{Gokhale} {et~al.}(2007){Gokhale}, {Peng}, \&
  {Frank}}]{2007ApJ...655.1010G}
{Gokhale}, V., {Peng}, X.~M., \& {Frank}, J. 2007, \apj, 655, 1010

\bibitem[{{Habets} \& {Zwaan}(1989)}]{1989A&A...211...56H}
{Habets}, G.~M.~H.~J. \& {Zwaan}, C. 1989, \aap, 211, 56

\bibitem[{{Hadjidemetriou}(1969{\natexlab{a}})}]{1969Ap&SS...3..330H}
{Hadjidemetriou}, J.~D. 1969{\natexlab{a}}, \apss, 3, 330

\bibitem[{{Hadjidemetriou}(1969{\natexlab{b}})}]{1969Ap&SS...3...31H}
{Hadjidemetriou}, J.~D. 1969{\natexlab{b}}, \apss, 3, 31

\bibitem[{{Han} {et~al.}(2002){Han}, {Podsiadlowski}, {Maxted}, {Marsh}, \&
  {Ivanova}}]{2002MNRAS.336..449H}
{Han}, Z., {Podsiadlowski}, P., {Maxted}, P.~F.~L., {Marsh}, T.~R., \&
  {Ivanova}, N. 2002, \mnras, 336, 449

\bibitem[{{Hoyle} \& {Fowler}(1960)}]{1960ApJ...132..565H}
{Hoyle}, F. \& {Fowler}, W.~A. 1960, \apj, 132, 565

\bibitem[{{Khaliullin} \& {Khaliullina}(2007)}]{2007MNRAS.382..356K}
{Khaliullin}, K.~F. \& {Khaliullina}, A.~I. 2007, \mnras, 382, 356

\bibitem[{{Khaliullin} \& {Khaliullina}(2010)}]{2010MNRAS.401..257K}
{Khaliullin}, K.~F. \& {Khaliullina}, A.~I. 2010, \mnras, 401, 257

\bibitem[{{Kolb} \& {Ritter}(1990)}]{1990A&A...236..385K}
{Kolb}, U. \& {Ritter}, H. 1990, \aap, 236, 385

\bibitem[{{Kruszewski}(1964{\natexlab{a}})}]{1964AcA....14..231K}
{Kruszewski}, A. 1964{\natexlab{a}}, \actaa, 14, 231

\bibitem[{{Kruszewski}(1964{\natexlab{b}})}]{1964AcA....14..241K}
{Kruszewski}, A. 1964{\natexlab{b}}, \actaa, 14, 241

\bibitem[{{Lehmann} {et~al.}(2013){Lehmann}, {Southworth}, {Tkachenko}, \&
  {Pavlovski}}]{2013A&A...557A..79L}
{Lehmann}, H., {Southworth}, J., {Tkachenko}, A., \& {Pavlovski}, K. 2013,
  \aap, 557, A79

\bibitem[{{Leigh} {et~al.}(2013){Leigh}, {Knigge}, {Sills}, {Perets},
  {Sarajedini}, \& {Glebbeek}}]{2013MNRAS.428..897L}
{Leigh}, N., {Knigge}, C., {Sills}, A., {et~al.} 2013, \mnras, 428, 897

\bibitem[{{Limber}(1963)}]{1963ApJ...138.1112L}
{Limber}, D.~N. 1963, \apj, 138, 1112

\bibitem[{{Lubow} \& {Shu}(1975)}]{1975ApJ...198..383L}
{Lubow}, S.~H. \& {Shu}, F.~H. 1975, \apj, 198, 383

\bibitem[{{Luk'yanov}(2008)}]{2008ARep...52..680L}
{Luk'yanov}, L.~G. 2008, Astronomy Reports, 52, 680

\bibitem[{{Maeder}(2009)}]{2009pfer.book.....M}
{Maeder}, A. 2009, {Physics, Formation and Evolution of Rotating Stars}

\bibitem[{{McCrea}(1964)}]{1964MNRAS.128..147M}
{McCrea}, W.~H. 1964, \mnras, 128, 147

\bibitem[{{Meibom} \& {Mathieu}(2005)}]{2005ApJ...620..970M}
{Meibom}, S. \& {Mathieu}, R.~D. 2005, \apj, 620, 970

\bibitem[{{Meibom} {et~al.}(2006){Meibom}, {Mathieu}, \&
  {Stassun}}]{2006ApJ...653..621M}
{Meibom}, S., {Mathieu}, R.~D., \& {Stassun}, K.~G. 2006, \apj, 653, 621

\bibitem[{{Mengel} {et~al.}(1976){Mengel}, {Norris}, \&
  {Gross}}]{1976ApJ...204..488M}
{Mengel}, J.~G., {Norris}, J., \& {Gross}, P.~G. 1976, \apj, 204, 488

\bibitem[{{Neo} {et~al.}(1977){Neo}, {Miyaji}, {Nomoto}, \&
  {Sugimoto}}]{1977PASJ...29..249N}
{Neo}, S., {Miyaji}, S., {Nomoto}, K., \& {Sugimoto}, D. 1977, \pasj, 29, 249

\bibitem[{{Packet}(1981)}]{1981A&A...102...17P}
{Packet}, W. 1981, \aap, 102, 17

\bibitem[{{Paczy{\'n}ski}(1971)}]{1971ARA&A...9..183P}
{Paczy{\'n}ski}, B. 1971, \araa, 9, 183

\bibitem[{{Paczynski}(1976)}]{1976IAUS...73...75P}
{Paczynski}, B. 1976, in IAU Symp. 73: Structure and Evolution of Close Binary
  Systems, ed. P.~{Eggleton}, S.~{Mitton}, \& J.~{Whelan}, 75

\bibitem[{{Piotrowski}(1964)}]{1964AcA....14..251P}
{Piotrowski}, S.~L. 1964, \actaa, 14, 251

\bibitem[{{Piotrowski}(1967)}]{1967SvA....11..191P}
{Piotrowski}, S.~L. 1967, \sovast, 11, 191

\bibitem[{{Pratt} \& {Strittmatter}(1976)}]{1976ApJ...204L..29P}
{Pratt}, J.~P. \& {Strittmatter}, P.~A. 1976, \apjl, 204, L29

\bibitem[{{Pringle} \& {Wade}(1985)}]{1985ibs..book.....P}
{Pringle}, J.~E. \& {Wade}, R.~A. 1985, {Interacting binary stars}

\bibitem[{{Qian} {et~al.}(2007){Qian}, {Yuan}, {Liu}, {He}, {Fern{\'a}ndez
  Laj{\'u}s}, \& {Kreiner}}]{2007MNRAS.380.1599Q}
{Qian}, S.-B., {Yuan}, J.-Z., {Liu}, L., {et~al.} 2007, \mnras, 380, 1599

\bibitem[{{Raymer}(2012)}]{2012MNRAS.427.1702R}
{Raymer}, E. 2012, \mnras, 427, 1702

\bibitem[{{Rieutord}(1992)}]{1992A&A...259..581R}
{Rieutord}, M. 1992, \aap, 259, 581

\bibitem[{{Rieutord} \& {Zahn}(1997)}]{1997ApJ...474..760R}
{Rieutord}, M. \& {Zahn}, J.-P. 1997, \apj, 474, 760

\bibitem[{{Savonije}(1978)}]{1978A&A....62..317S}
{Savonije}, G.~J. 1978, \aap, 62, 317

\bibitem[{{Sepinsky} {et~al.}(2007{\natexlab{a}}){Sepinsky}, {Willems}, \&
  {Kalogera}}]{2007ApJ...660.1624S}
{Sepinsky}, J.~F., {Willems}, B., \& {Kalogera}, V. 2007{\natexlab{a}}, \apj,
  660, 1624

\bibitem[{{Sepinsky} {et~al.}(2007{\natexlab{b}}){Sepinsky}, {Willems},
  {Kalogera}, \& {Rasio}}]{2007ApJ...667.1170S}
{Sepinsky}, J.~F., {Willems}, B., {Kalogera}, V., \& {Rasio}, F.~A.
  2007{\natexlab{b}}, \apj, 667, 1170

\bibitem[{{Sepinsky} {et~al.}(2010){Sepinsky}, {Willems}, {Kalogera}, \&
  {Rasio}}]{2010ApJ...724..546S}
{Sepinsky}, J.~F., {Willems}, B., {Kalogera}, V., \& {Rasio}, F.~A. 2010, \apj,
  724, 546

\bibitem[{{Siess}(2010)}]{2010A&A...512A..10S}
{Siess}, L. 2010, \aap, 512, A10

\bibitem[{{Siess} {et~al.}(2013){Siess}, {Izzard}, {Davis}, \&
  {Deschamps}}]{2013A&A...550A.100S}
{Siess}, L., {Izzard}, R.~G., {Davis}, P.~J., \& {Deschamps}, R. 2013, \aap,
  550, A100

\bibitem[{{Song} {et~al.}(2013){Song}, {Maeder}, {Meynet}, {Huang},
  {Ekstr{\"o}m}, \& {Granada}}]{2013A&A...556A.100S}
{Song}, H.~F., {Maeder}, A., {Meynet}, G., {et~al.} 2013, \aap, 556, A100

\bibitem[{{St{\c e}pie{\'n}}(2009)}]{2009MNRAS.397..857S}
{St{\c e}pie{\'n}}, K. 2009, \mnras, 397, 857

\bibitem[{{Sterne}(1960)}]{1960aitc.book.....S}
{Sterne}, T.~E. 1960, {An introduction to celestial mechanics}

\bibitem[{{Sudar} {et~al.}(2011){Sudar}, {Harmanec}, {Lehmann}, {Yang}, {Bo{\v
  z}i{\'c}}, \& {Ru{\v z}djak}}]{2011A&A...528A.146S}
{Sudar}, D., {Harmanec}, P., {Lehmann}, H., {et~al.} 2011, \aap, 528, A146

\bibitem[{{S{\"u}rgit} {et~al.}(2010){S{\"u}rgit}, {Erdem}, \&
  {Budding}}]{2010MNRAS.407..497S}
{S{\"u}rgit}, D., {Erdem}, A., \& {Budding}, E. 2010, \mnras, 407, 497

\bibitem[{{Sytov} {et~al.}(2007){Sytov}, {Kaigorodov}, {Bisikalo}, {Kuznetsov},
  \& {Boyarchuk}}]{2007ARep...51..836S}
{Sytov}, A.~Y., {Kaigorodov}, P.~V., {Bisikalo}, D.~V., {Kuznetsov}, O.~A., \&
  {Boyarchuk}, A.~A. 2007, Astronomy Reports, 51, 836

\bibitem[{{Tassoul}(1987)}]{1987ApJ...322..856T}
{Tassoul}, J.-L. 1987, \apj, 322, 856

\bibitem[{{Tassoul}(1988)}]{1988ApJ...324L..71T}
{Tassoul}, J.-L. 1988, \apjl, 324, L71

\bibitem[{{Tassoul} \& {Tassoul}(1997)}]{1997ApJ...481..363T}
{Tassoul}, M. \& {Tassoul}, J.-L. 1997, \apj, 481, 363

\bibitem[{{Tout} \& {Eggleton}(1988{\natexlab{a}})}]{1988ApJ...334..357T}
{Tout}, C.~A. \& {Eggleton}, P.~P. 1988{\natexlab{a}}, \apj, 334, 357

\bibitem[{{Tout} \& {Eggleton}(1988{\natexlab{b}})}]{1988MNRAS.231..823T}
{Tout}, C.~A. \& {Eggleton}, P.~P. 1988{\natexlab{b}}, \mnras, 231, 823

\bibitem[{{Ulrich} \& {Burger}(1976)}]{1976ApJ...206..509U}
{Ulrich}, R.~K. \& {Burger}, H.~L. 1976, \apj, 206, 509

\bibitem[{{van Rensbergen} {et~al.}(2008){van Rensbergen}, {De Greve}, {De
  Loore}, \& {Mennekens}}]{2008A&A...487.1129V}
{van Rensbergen}, W., {De Greve}, J.~P., {De Loore}, C., \& {Mennekens}, N.
  2008, \aap, 487, 1129

\bibitem[{{Vanbeveren}(1977)}]{1977A&A....54..877V}
{Vanbeveren}, D. 1977, \aap, 54, 877

\bibitem[{{Wang} \& {Han}(2012)}]{2012NewAR..56..122W}
{Wang}, B. \& {Han}, Z. 2012, \nar, 56, 122

\bibitem[{{Webbink}(1977{\natexlab{a}})}]{1977ApJ...211..486W}
{Webbink}, R.~F. 1977{\natexlab{a}}, \apj, 211, 486

\bibitem[{{Webbink}(1977{\natexlab{b}})}]{1977ApJ...215..851W}
{Webbink}, R.~F. 1977{\natexlab{b}}, \apj, 215, 851

\bibitem[{{Webbink}(2008)}]{2008ASSL..352..233W}
{Webbink}, R.~F. 2008, in Astrophysics and Space Science Library, Vol. 352,
  Astrophysics and Space Science Library, ed. E.~F. {Milone}, D.~A. {Leahy}, \&
  D.~W. {Hobill}, 233

\bibitem[{{Yakut} {et~al.}(2007){Yakut}, {Aerts}, \&
  {Morel}}]{2007A&A...467..647Y}
{Yakut}, K., {Aerts}, C., \& {Morel}, T. 2007, \aap, 467, 647

\bibitem[{{Zahn}(1977)}]{1977A&A....57..383Z}
{Zahn}, J.-P. 1977, \aap, 57, 383

\bibitem[{{Zahn}(1989)}]{1989A&A...220..112Z}
{Zahn}, J.-P. 1989, \aap, 220, 112

\bibitem[{{Zeldovich} \& {Guseynov}(1966)}]{1966ApJ...144..840Z}
{Zeldovich}, Y.~B. \& {Guseynov}, O.~H. 1966, \apj, 144, 840

\bibitem[{{Zhao} {et~al.}(2014){Zhao}, {Qian}, {Li}, {He}, {Liu}, {Wang}, \&
  {Zhang}}]{2014NewA...26..112Z}
{Zhao}, E., {Qian}, S., {Li}, L., {et~al.} 2014, \na, 26, 112

\end{thebibliography}

\end{document}